\documentclass[runningheads]{llncs}
\usepackage{graphicx}
\usepackage{microtype}

\usepackage[shortlabels]{enumitem}


\usepackage{subcaption}

\usepackage{makecell}
\usepackage{xcolor}
\usepackage{hyperref}
\hypersetup{backref=true,
    pdftex=true,
    pagebackref=true,               
    hyperindex=true,                
    colorlinks=true,                
    breaklinks=true,                
    urlcolor= black,                
    linkcolor= blue,                
    bookmarks=true,                 
    bookmarksopen=false,
    filecolor=black,
    citecolor=blue,
    linkbordercolor=blue
}
\AtBeginDocument{\hypersetup{pdfborder={0 0 1}}}
\usepackage{hypcap}

\usepackage[noend]{algpseudocode}
\algdef{SE}[DOWHILE]{Do}{doWhile}{\algorithmicdo}[1]{\algorithmicwhile\ #1}%

\algnewcommand{\IIf}[1]{\State\algorithmicif\ #1\ \algorithmicthen}
\algnewcommand{\EndIIf}{\unskip\ \algorithmicend\ \algorithmicif}
\algdef{SE}[DOWHILE]{Do}{doWhile}{\algorithmicdo}[1]{\algorithmicwhile\ #1}%

\usepackage{amssymb}
\usepackage{amsmath,wasysym}
\usepackage{cite}

\usepackage[title]{appendix}

\begin{document}

\title{Revisited Containment for Graph Patterns}
\titlerunning{Revisited Containment for Graph Patterns}
%
\author{Houari Mahfoud\orcidID{0000-0003-0277-1928}}
\authorrunning{H. Mahfoud}
%
\institute{Abou-Bekr Belkaid University \& LRIT Laboratory, Tlemcen, Algeria\\
\email{houari.mahfoud@\{univ-tlemcen.dz, gmail.com\}}}
\maketitle              

\begin{abstract}
We consider the class of conditional graph patterns (\emph{CGPs}) that allow user to query data graphs with complex patterns that contain negation and predicates. 
To overcome the prohibitive cost of subgraph isomorphism, we consider matching of \emph{CGPs} under simulation semantics which can be conducted in quadratic time. 
In emerging applications, one would like to reduce more this matching time, and the static analysis of patterns may allow ensuring part of this reduction. 
We study the containment problem of \emph{CGPs} that aims to check whether the matches of some pattern $P_1$, over any data graph, are contained in those of another pattern $P_2$ (written $P_1\sqsubseteq P_2$). The optimization process consists to extract matches of $P_1$ only from those of $P_2$ without querying the (possibly large) data graph. We show that the traditional semantics of containment is decidable in quadratic time, but it fails to meet the optimization goal in the presence of negation and predicates. To overcome this limit, we propose a new semantics of containment, called \emph{strong containment}, that is more suitable for \emph{CGPs} and allows to reduce their matching time. We show that \emph{strong containment} can be decided in cubic time by providing such an algorithm. We are planing to use results of this paper to answer \emph{CGPs} using views. This paper is part of an ongoing project that aims to design a caching system for complex graph patterns.

\end{abstract}

\section{Introduction}
Given a data graph $G$ and a graph pattern $P$, graph pattern matching (\emph{GPM}) is to find all subgraphs of $G$ that match $P$. Matching is traditionally expressed in terms of subgraph isomorphism which is cost prohibitive. To avoid this cost, graph simulation \cite{GraphSimulation} and its extensions  \cite{GPM_From_Intractable_To_Polynomial_Time,Adding_Reg_Exp_Fan_11,Fan14} have been proposed that allow \emph{GPM} to be approximately conducted in polynomial time. However, existing simulation-based \emph{GPM} consider very simple patterns which do not meet requirements of real-life applications. To overcome this limit, we proposed in \cite{Mahfoud20} conditional graph patterns (\emph{CGPs}) which allow to query data graphs with complex features like quantifications, predicates and negation.
Our goal is to study the containment problem of \emph{CGPs} in order to reduce their matching time. Given two \emph{CGPs} $C_1$ and $C_2$, it is to decide whether the matches of $C_1$ over any data graph are all contained in those of $C_2$. If such is the case, then one can reduce matching time of $C_1$ by extracting its matches from those of $C_2$. The containment problem has been widely studied for different query languages (e.g. relational queries \cite{Relational_Containment_1}, XPath queries \cite{XPath_Containment_2}, reachability queries \cite{Containment_of_RGQs}). When it comes to graph patterns however, the containment problem has not received sufficient attention since it has been studied by only few works that have considered only simple patterns (e.g. \cite{Adding_Reg_Exp_Fan_11}). Moreover, its decision problem has been studied under a traditional semantics which does not apply for complex patterns and also hinders the realization of some tasks. We explain these limits by the next example.

\begin{figure}[t!]
\centering
   \noindent\makebox[\textwidth]{%
   \includegraphics[width=1.1\linewidth,height=4cm]{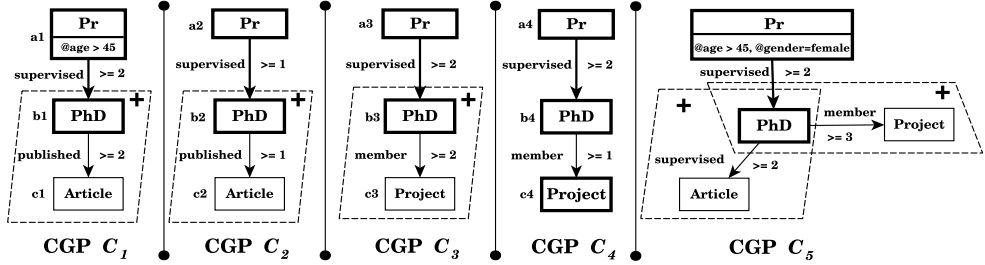}}
   \caption{Example of expressive graph patterns.}
   \label{figure:Introduction_Example}
\end{figure}

\begin{example}\label{example:introduction}
Consider the graph patterns $C_{i\in[1,5]}$ of Fig. \ref{figure:Introduction_Example}. Throughout this paper, part of a pattern in bold form represents the structure of the results that will be returned to the user, called the \emph{core} of the pattern, while the remaining parts represent positive and/or negative predicates used to impose some conditions over nodes of the core. Moreover, variables are attached to nodes of our patterns to simplify their referencing. $C_1$ looks for professors (with \emph{age} $\geq 45$) which supervised at least two PhD students providing that each student has published at least two articles. The articles published by these PhD students represent a positive condition so they will not returned to the user as result of $C_1$. In other words, when evaluating $C_1$ over any data graph, the results will be composed only by nodes, labeled by \emph{Pr} and \emph{PhD}, that are connected by some edges labeled \emph{supervised}. Remark that the results returned by $C_1$ (resp. $C_3$) will be contained in those returned by $C_2$ (resp. $C_4$) over any data graph $G$ since the former pattern is a special case of the latter one. We denote that by $C_1\sqsubseteq C_2$ (resp. $C_3\sqsubseteq C_4$). These cases can be checked using the well-known definition of containment. However, given the results of $C_2$ over $G$, one cannot refine them to find those of $C_1$ over $G$. This is due to the fact that the PhD students returned by $C_2$ have published at least one article, but these articles are not returned to the user as part of the results of $C_2$ over $G$, which makes impossible the refinement of these PhD students to keep only those who have published at least two articles, i.e. those that are supposed to be returned by $C_1$. On the other side, $C_4$ returns the PhD students with their corresponding projects so one can refine these PhD students to keep only those that are supposed to be returned by $C_3$. In other words, the traditional containment tells just whether the results of some pattern are contained in those of another one but it does not help to decide whether the extraction of these results is possible. This may hinder the answering of complex patterns using views. Remark that the results of $C_5$ over any data graph $G$ can be computed by merging those of $C_1$ and $C_4$ over $G$, but not those of $C_1$ and $C_3$. That is, $C_5\sqsubseteq C_1\cup C_4$ can be checked if one can decide whether the results of some part of $C_5$ can be extracted from those of $C_1$ (resp. $C_4$), which is not possible using the traditional containment.\hfill$\Square$
\end{example}

This tells us that complex patterns deserve the definition of a new semantics of containment in order to allow achieving more complex tasks with these patterns like query optimization and minimization, and views-based answering.\\

\noindent\textbf{Contributions and Road-map.} 
We first recall from \cite{Mahfoud20, Mahfoud21} the definition of \emph{CGPs} and we show how they can be matched in quadratic time using conditional simulation 
(Section \ref{section:cgps}). Next, we make the following contributions. (\textit{i}) We propose the notion of \emph{pattern-only matching} that checks whether a pattern 
satisfies all constraints of another one (Section \ref{section:pom}). This notion is necessary to decide containment of our patterns. In Section \ref{section:tc}, (\textit{ii}) 
we revise the formal definition of the traditional containment to be applied for \emph{CGPs}, and (\textit{iii}) we provide a quadratic-time algorithm to decide it. 
(\textit{iv}) We propose a new semantics of containment, called \emph{strong containment}, that applies well for patterns with predicates and negation, and (\textit{v}) 
we show that this semantics can be decided in cubic time by providing such an algorithm (Section \ref{section:sc}). Finally, we discuss how to match efficiently \emph{CGPs} using \emph{strong containment} (Section \ref{section:matching_via_sc}).\footnote{Due to space limitations, the proofs are given in Appendix.}\\

\noindent\textbf{Related Work.} The pattern containment is a classical and fundamental problem for any query language. It has been well studied for relational queries (e.g. \cite{Relational_Containment_1}) and XML queries (e.g., \cite{Containment_of_XPath,XPath_Containment_2}). 
For graph patterns however, it is striking how little attention has been paid for this problem. The containment problem has been studied in \cite{Adding_Reg_Exp_Fan_11} for graph patterns without neither predicates nor negation, which makes the task less intriguing.
Moreover, the problem has been studied in \cite{Containment_of_RGQs} for graph reachability queries with memory which is not closer to our work. Apart from these works, we are not aware of other ones that study the containment problem in case of complex patterns.

\section{Preliminaries}

\subsection{Data Graphs and Graph Patterns}

\noindent \textbf{Data Graphs.} 
A data graph is a directed graph $G$=($V,E,\mathcal{L},\mathcal{A}$) where: 
1) $V$ is a finite set of nodes; 2) $E\subseteq V\times V$ is a finite set of edges in which $(v,v^{'})$ denotes an edge from node $v$ to $v^{'}$; 3) $\mathcal{L}$ is a function that assigns a label $\mathcal{L}(v)$ (resp. $\mathcal{L}(e)$)  to each node $v\in V$ (resp. edge $e\in E$); and 4) for each node $v\in V$, $\mathcal{A}(v)$ is a tuple $(A_1=c_1,\dots,A_n=c_n)$ where: $A_i$ is an attribute of $v$, $c_i$ is a constant value, $n\geq 0$, and $A_i\neq A_j$ if $i\neq j$. We denote by $v.A_i$ the value of attribute $A_i$ on node $v$ (i.e. value $c_i$).

Intuitively, the label of a node represents an entity (e.g. \emph{Movie}, \emph{Person}) while the label of an edge represents a relationship (e.g. $Person\xrightarrow{producedBy}Movie$, $Person\xrightarrow{isFriend}Person$). Moreover, the function $\mathcal{A}$ defines properties over nodes (e.g. \emph{title} and \emph{release date} of movies).

\noindent \textbf{Subgraphs.}
\sloppy{Given a data graph $G$=($V,E,\mathcal{L},\mathcal{A}$), a subgraph $G_s$=($V_s,E_s,\mathcal{L}_s,\mathcal{A}_s$) of $G$=($V,E,\mathcal{L},\mathcal{A}$) must satisfy: 1) $V_s\!\subseteq\! V$; 2) $E_s\!\subseteq\! E$; 3) $\mathcal{L}_s(x)\!=\!\mathcal{L}(x)$ for each $x\in E_s\!\cup\!V_s$; and 4) $\mathcal{A}_s(v)\!=\!\mathcal{A}(v)$ for each $v\in V_s$.}

\noindent \textbf{Conventional Graphs Patterns.}
A graph pattern is a directed connected graph $P$=($V,E,\mathcal{L},\mathcal{A}$) where: 1) $V$, $E$, and $\mathcal{L}$ are defined as for data graphs; and 2) for each node $u\in V$, $\mathcal{A}(u)$ is a predicate defined as a conjunction of atomic formulas of the form ``$A$ $op$ $c$'' where: $A$ is an attribute of $u$, $c$ is a constant, and $op\in\{\geq, \leq, =, \neq\}$. Intuitively, $\mathcal{A}(u)$ specifies a search condition on a node $u$. 

\noindent\textbf{Paths.} An \emph{undirected path} $\textlangle e_1,\cdots,e_n\textrangle$ in $P$ is a sequence of edges of $E$ where: for each $1\leq i\leq n$, $e_i$ intersects with $e_{i+1}$ in some node. For instance, $\textlangle e_1,e_2,e_3\textrangle$ with $e_1=a\rightarrow b$, $e_2=c\rightarrow b$ and $e_3=d\rightarrow c$ is an undirected path.


\subsection{Conventional Graph Pattern Matching}\label{section:conventional_GPM}
We refer hereafter to the data graph $G$=($V,E,\mathcal{L},\mathcal{A}$) and the conventional graph pattern $P$=($V_{_P},E_{_P},\mathcal{L}_{_P},\mathcal{A}_{_P}$). 
We next revise definitions of these three graph pattern matching models.

\noindent\textbf{Attribute Constraints Matching.}
Given a pattern node $u\in V_P$ and a data node $v\in V$, we say that attributes values of $v$ match attributes constraints of $u$, written $\mathcal{A}(v)\sim\mathcal{A}_P(u)$, if and only if: for each atomic formula ``$A~op~c$'' in $\mathcal{A}_{_P}(u)$, there exists $A=c^{'}$ in $\mathcal{A}(v)$ where $c^{'}~op~c$ holds.

\noindent\textbf{Subgraph Isomorphism.}
A subgraph $G_s$=($V_s,E_s,\mathcal{L}_s,\mathcal{A}_s$) of $G$ matches $P$ via \emph{subgraph isomorphism} if there exists a \emph{bijective function} $f$:$V_{_P}\rightarrow V_s$ s.t.: \textit{1}) for each $u\in V_{_P}$, $\mathcal{L}_{_P}(u)=\mathcal{L}_s(f(u))$ and $\mathcal{A}_s(f(u))\sim\mathcal{A}_{_P}(u)$; \textit{2}) for each $e_u=(u,u^{'})$ in $E_P$, there exists an edge $e_s=(f(u),f(u^{'}))$ in $G_s$ with $\mathcal{L}_{_P}(e_u)=\mathcal{L}_s(e_s)$.

\noindent\textbf{Graph Simulation.}
$G$ matches $P$ via \emph{graph simulation} if there exists a \emph{binary match relation} $S\subseteq V_{_P}\times V$ s.t.: 
\textit{1}) for each $(u,v)\in S$, $\mathcal{L}_{_P}(u)=\mathcal{L}(v)$ and $\mathcal{A}(v)\sim\mathcal{A}_{_P}(u)$; \textit{2}) for each $(u,v)\in S$ and each edge $e_u=(u,u^{'})$ in $E_{_P}$, there exists an edge $e_v=(v,v^{'})$ in $E$ with $(u^{'},v^{'})\in S$ and $\mathcal{L}_{_P}(e_u)=\mathcal{L}(e_v)$; and \textit{3}) for each $u\in V_{_P}$, there is at least one node $v\in V$ with $(u,v)\in S$.

By condition (2), graph simulation preserves only child relationships.

\noindent\textbf{Dual Simulation.}
$G$ matches $P$ via \emph{dual simulation} if there exists a \emph{binary match relation} $S_D\subseteq V_{_P}\times V$ s.t.:
\textit{1}) for each $(u,v)\in S$, $\mathcal{L}_{_P}(u)=\mathcal{L}(v)$ and $\mathcal{A}(v)\sim\mathcal{A}_{_P}(u)$; \textit{2}) for each $(u,v)\in S$ and each edge $e_u=(u,u^{'})$ (resp. $e_u=(u^{'},u)$) in $E_{_P}$, there exists an edge $e_v=(v,v^{'})$ (resp. $e_v=(v^{'},v)$) in $E$ with $(u^{'},v^{'})\in S$ and $\mathcal{L}_{_P}(e_u)=\mathcal{L}(e_v)$; and \textit{3}) for each node $u\in V_{_P}$, there exists at least one node $v\in V$ with $(u,v)\in S$.\footnote{By condition (2), dual simulation preserves both child and parent relationships.}

\section{Conditional Graph Patterns (\emph{CGPs})}\label{section:cgps}
\subsection{Definition of \emph{CGPs}}\label{subsection:definition_of_cgps}
We start by extending conventional patterns with simple \emph{counting quantifiers (CQs)}, which leads to \emph{quantified graph patterns (QGPs)}.

\begin{definition}\label{section:syntax_of_quantified_graph_patterns} 
A \emph{QGP} is a connected directed graph $Q$=($V,E,\mathcal{L},\mathcal{A},\mathcal{C}$) where: 1) $V$, $E$, $\mathcal{L}$, $\mathcal{A}$ are 
defined as for conventional graph patterns; and 2) for each edge $e\in E$, $\mathcal{C}(e)$ is a \emph{CQ} given by an integer $p$ ($p\geq 1$).\hfill$\Square$
\end{definition}

Intuitively, for any data graph $G$ and any edge $e=(u,u')$ in $Q$ with a \emph{CQ} $\mathcal{C}(e)=p$, a data node $v$ in $G$ matches $u$ if it has at least $p$ children that match $u'$, and moreover, these children must be reached from $v$ via an edge labeled $\mathcal{L}(e)$. We omit \emph{CQs} that are equal to $1$ for more readability. Matching $Q$ over $G$ consists to find all subgraphs of $G$ that match the structure (i.e. nodes and edges relationships) and constraints (i.e. labeling, attributes and \emph{CQs}) of $Q$. We denote this match result by $\mathcal{M}^{G}_{_Q}$ and we report its definition to the next subsection.

Along the same lines as \cite{CQ_Fan_16}, we define a special form of \emph{QGPs} by considering the \emph{query focus}.

\begin{definition}\label{section:syntax_of_quantified_graph_patterns}
We denote by $Q(u)$ a QGP $Q$=($V,E,\mathcal{L},\mathcal{A},\mathcal{C}$) with a \emph{query focus} $u\in V$ used for search intent. Indeed, the result of $Q(u)$ over any data graph $G$ is a nodes set, extracted from $\mathcal{M}^{G}_{_Q}$, that contains all data nodes in $G$ that match $u$.\hfill$\Square$
\end{definition}

Using \emph{QGPs} (general and special forms) as building blocks, we next define a new class of graph patterns called \emph{conditional graph patterns (CGPs)}.

\begin{definition}\label{definition:CGPs} 
A \emph{CGP} is a connected directed graph $C$=($V,E,\mathcal{L},\mathcal{A},\mathcal{C},\mathcal{P}^{+},\mathcal{P}^{-}$) where:

\begin{enumerate}
 \item ($V,E,\mathcal{L},\mathcal{A},\mathcal{C}$) is a QGP called the \emph{core} of $C$; and
 
 \item $\mathcal{P}^{+}$ (resp. $\mathcal{P}^{-}$) is a set of positive (resp. negative) predicates defined over nodes in $V$ such that:
 
 \begin{enumerate}
  \item each $p^{+}(u)\in \mathcal{P}^{+}$ (resp. $p^{-}(u)\in \mathcal{P}^{-}$) is a QGP that defines a positive (resp. negative) predicate with the query focus $u\in V$; and
  \item $p^{+}(u)$ (resp. $p^{-}(u)$) intersects with $V$ only on the node $u$.\footnote{This may not reduce the practicability of our approach since many query languages (e.g. XPath, SQL) adopt this syntax of predicates.}\hfill$\Square$
 \end{enumerate}
\end{enumerate}
\end{definition}

Remark that \emph{CGPs} extend \emph{QGPs} by incorporating two (possibly empty) sets of positive and negative predicates, $\mathcal{P}^{+}$ and $\mathcal{P}^{-}$. Contrary to conventional patterns where conditions are expressed only in terms of attributes, our syntax allows the definition of a \emph{quantified and attributed graph-based condition}. A \emph{core node} (resp. \emph{core edge}) is any node (resp. edge) that belongs to the core of $C$, i.e. the set $V$ (resp. $E$). Moreover, a \emph{predicate node} (resp. \emph{predicate edge}) is any node (resp. edge) that belongs to some predicate in $\mathcal{P}^{+}\cup\mathcal{P}^{-}$.

The semantic of a \emph{CGP} $C$ is stated as follows. A data graph $G$ matches $C$ if and only if: 1) it has a subgraph $G_s$ that matches the core of $C$; and 2) for any node $u\in V$ and any predicate $p^{+}(u)\in \mathcal{P}^{+}$ (resp. $p^{-}(u)\in \mathcal{P}^{-}$), a node $v$ in $G_s$ is a match of $u$ if it belongs (resp. does not belong) to the nodes set returned by $p^{+}(u)$ (resp. $p^{-}(u)$) over $G$.


Inspired by well-known conditional languages (e.g. SQL, XPath), the core of a \emph{CGP} $C$ represents the structure of the match result that will be returned to the user, while predicates in $\mathcal{P}^{+}\cup \mathcal{P}^{-}$ are used only during the matching process to refine this result. To our knowledge, the expressivity given by our \emph{CGPs} is not covered by any approach in the literature  (e.g \cite{Fan14,Adding_Reg_Exp_Fan_11,CQ_Fan_16,GPM_From_Intractable_To_Polynomial_Time}).

\begin{example}\label{example:cgps}
Consider the patterns $C_{i\in[1,5]}$ depicted in Fig. \ref{figure:Introduction_Example} where $C_{4}$ is a \emph{QGP}, since it contains no predicate, while the other ones are \emph{CGPs}. The core of $C_2$ is given by ${\scriptstyle Pr\xrightarrow[\geq 2]{supervised}PhD}$ and represents the structure of the matches that will be returned by $C_2$ over data graphs. The remaining part of $C_2$, i.e. ${\scriptstyle PhD\xrightarrow[\geq 1]{published}Article}$, is a positive predicate used to refine the PhD students returned by $C_2$. Remark that each predicate intersects with the core in exactly one node, and there may be different predicates over the same node (case of $C_5$).\hfill$\Square$
\end{example}

\begin{definition}\label{definition:positive_version_of_CGP}
Given a CGP $C$=($V,E,\mathcal{L},\mathcal{A},\mathcal{C},\mathcal{P}^{+},\mathcal{P}^{-}$). We denote by $V^{+}$ (resp. $E^{+}$) the set of all nodes (resp. edges) composing positive predicates in $\mathcal{P}^{+}$. Similarly for $V^{-}$ and $E^{-}$. 
The \emph{positive version} of $C$, denoted by $C^{+}$, is a QGP composed by the nodes set $V\cup V^{+}$, the edges set $E\cup E^{+}$, and their corresponding labels, attributes and CQs defined in $C$.\hfill$\Square$
\end{definition}

Given the \emph{CGP} $C_2$ of Example \ref{example:cgps}, then $C^{+}_2$ is given 
by ${\scriptstyle Pr\xrightarrow[\geq 2]{supervised}PhD\xrightarrow[\geq 1]{published} Article}$.

For any \emph{QGP} $Q$ with nodes set $V_Q$ and edges set $E_Q$, then the size of $Q$ (i.e. $|Q|$) is given by $|V_Q|+|E_Q|$. For any CGP $C$=($V,E,\mathcal{L},\mathcal{A},\mathcal{C},\mathcal{P}^{+},\mathcal{P}^{-}$), the size of $C$ (i.e. $|C|$) is given by $|V|+|E|+|\mathcal{P}^{+}|+|\mathcal{P}^{-}|$ where $|\mathcal{P}^{+}|$ (resp. $|\mathcal{P}^{-}|$) is the total number of nodes and edges composing positive (resp. negative) predicates in $C$. Moreover, $|C|=|C^{+}|+|\mathcal{P}^{-}|$.

\subsection{Conditional Graph Pattern Matching}\label{subsection:matching_cgps}
In order to overcomes the prohibitive cost of isomorphism, we propose to match \emph{CGPs} via an extension of graph simulation, called \emph{conditional simulation}.

\begin{definition}\label{definition:conditional_simulation}
\sloppy{A data graph $G=(V,E,\mathcal{L},\mathcal{A})$ matches a \textit{CGP} $C=(V_{_C},E_{_C},\mathcal{L}_{_C},\mathcal{A}_{_C},\mathcal{C},\mathcal{P}^{+},\mathcal{P}^{-})$ via \emph{conditional simulation}, denoted by $C\prec_{_c} G$, if there exists a binary match relation $S^{G}_{_C}\subseteq V_{_C} \times V$ s.t.:}

\begin{enumerate}
    \item For each $(u,v)\in S^{G}_{_C}$: $\mathcal{L}_{_C}(u)$=$\mathcal{L}(v)$ and $\mathcal{A}(v)\sim\mathcal{A}_{_C}(u)$.
    
    \item For each $(u,v)\in S^{G}_{_C}$ and each $e_u=(u,u^{'})\in E_{_C}$ with $\mathcal{C}(e_u)$=$n$, there are at least $n$ edges $e_1$=$(v,v_1),\dots,e_n$=$(v,v_n)$ in $E$ s.t: $\mathcal{L}(e_i)$=$\mathcal{L}_{_C}(e_u)$ and $(u^{'},v_i)\in S^{G}_{_C}$ for $i\in[1,n]$.
    
    \item For each $(u,v)\in S^{G}_{_C}$ and each edge $e_u=(u^{'},u)$ in $E_{_C}$, there is at least one edge $e_v=(v^{'},v)$ in $E$ s.t: $\mathcal{L}(e_v)=\mathcal{L}_{_C}(e_u)$ and $(u^{'},v^{'})\in S^{G}_{_C}$.
    
    \item For each $(u,v)\in S^{G}_{_C}$ and each positive predicate $p^{+}(u)\in\mathcal{P}^{+}$, there is a subgraph $G_s\subseteq G$ s.t.: $p^{+}(u)\prec_{_c} G_s$ with a match relation $S$; and $(u,v)\in S$.
    
    \item For each $(u,v)\in S^{G}_{_C}$ and each $p^{-}(u)\in\mathcal{P}^{-}$, there is no subgraph $G_s\subseteq G$ s.t.: $p^{-}(u)\prec_{_c} G_s$ with a match relation $S$; and $(u,v)\in S$.
    
    \item Each node $u\!\in\!V_{_C}$ has at least one match $(u,v)\!\in\!S^{G}_{_C}$.\hfill$\Square$
\end{enumerate}
\end{definition}

Conditional simulation extends dual simulation by condition (2), in order to consider simple \emph{CQs} on core edges of $C$; and by conditions (4-5) to deal with predicates defined over core nodes of $C$. Since a predicate $p^{+}(u)$ (resp. $p^{-}(u)$) is a \emph{QGP}, i.e. a \emph{CGP} with no predicate, then one can check whether $p^{+}(u)\prec_{_c} G_s$ (resp. $p^{-}(u)\prec_{_c} G_s$) by considering only conditions (1--3,6) of Def. \ref{definition:conditional_simulation}.

When $C\prec_{_c} G$, there exists a unique maximum match relation $S^{G}_{_C}$ in
$G$ for $C$ \cite{Mahfoud20}. We derive from $S^{G}_{_C}$ the function $\mathcal{M}^{G}_{_C}$, called the \emph{match result} of $C$ in $G$, where: for any core node $u\in V_{_C}$, $\mathcal{M}^{G}_{_C}(u)=\{v\in V\backslash (u,v)\in S^{G}_{_C}\}$; and moreover, for any core edge $e_u=(u,u')\in E_{_C}$, $\mathcal{M}^{G}_{_C}(e_u)=\{e_v=(v,v')\in E\backslash \{(u,v),(u',v')\}\in S^{G}_{_C}, ~and~ \mathcal{L}(e_v)=\mathcal{L}_{_C}(e_u)\}$.

We have shown in \cite{Mahfoud20} that the checking of $C\prec_{_c} G$ as well as the computation of the maximum match relation $S^{G}_{_C}$ can be done in quadratic time. We proposed in \cite{Mahfoud21} two techniques to reduce matching time of \emph{CGPs}.

\section{Pattern-Only Matching}\label{section:pom}
We introduce the notion of \emph{pattern-only matching} that aims to check, for two graph patterns $Q_1$ and $Q_2$, whether $Q_1$ matches all constraints of $Q_2$.

\begin{definition}
Given two QGPs $Q_{i\in[1,2]}=(V_{_i},E_{_i},\mathcal{L}_{_i},\mathcal{A}_{_i},\mathcal{C}_{_i})$. I) For any two pattern nodes $u_1\in V_{_1}$ and $u_2\in V_{_2}$, we say that $\mathcal{A}_{_1}(u_1)$ \emph{matches} $\mathcal{A}_{_2}(u_2)$, written $\mathcal{A}_{_1}(u_1)\sim\mathcal{A}_{_2}(u_2)$, if the constraints defined over any attribute $A$ in $\mathcal{A}_{_2}(u_2)$ are satisfied by those defined over $A$ in $\mathcal{A}_{_1}(u_1)$. II) We say that $u_1$ \emph{matches} $u_2$ (i.e. $u_1\sim u_2$) if: $\mathcal{L}_{_1}(u_1)=\mathcal{L}_{_2}(u_2)$ and $\mathcal{A}_{_1}(u_1)\sim\mathcal{A}_{_2}(u_2)$. Moreover, III) for any two edges $e_1=(u_1,w_1)\in E_{_1}$ and $e_2=(u_2,w_2)\in E_{_2}$, we say that $e_1$ \emph{matches} $e_2$ (i.e. $e_1\sim e_2$) if: $u_1\sim u_2$; $w_1\sim w_2$; $\mathcal{L}_{_1}(e_1)=\mathcal{L}_{_2}(e_2)$ and $\mathcal{C}_{_1}(e_1)\geq\mathcal{C}_{_2}(e_2)$.\hfill$\Square$
\end{definition}

For instance, if we have $\mathcal{A}_{_1}(u_1)=``age>25,gender=femal''$ and $\mathcal{A}_{_2}(u_2)=``age\neq 20''$, then it is clear that $\mathcal{A}_{_1}(u_1)\sim \mathcal{A}_{_2}(u_2)$ but $\mathcal{A}_{_2}(u_2)\not\sim \mathcal{A}_{_1}(u_1)$.

\begin{definition}\label{definition:quantified_pattern_only_matching}
Given two QGPs $Q_{i\in[1,2]}=(V_{_i},E_{_i},\mathcal{L}_{_i},\mathcal{A}_{_i},\mathcal{C}_{_i})$. We say that $Q_1$ \emph{matches} $Q_2$, denoted by $Q_1\vartriangleright Q_2$, if there exists a binary match relation $S\subseteq V_{_1}\times V_{_2}$ such that:

\begin{enumerate}
  \item for each $(u_1,u_2)\in S$: $\mathcal{L}_{_1}(u_1)=\mathcal{L}_{_2}(u_2)$ and $\mathcal{A}_{_1}(u_1)\sim\mathcal{A}_{_2}(u_2)$.

  \item for each $(u_1,u_2)\in S$ and each edge $e_2=(w_2,u_2)$ in $E_{_2}$, there exists an edge $e_1=(w_1,u_1)$ in $E_{_1}$ with: $(w_1,w_2)\in S$ and $\mathcal{L}_{_1}(e_1)=\mathcal{L}_{_2}(e_2)$.
  
  \item for each $(u_1,u_2)\in S$ and each edge $e_2=(u_2,w_2)\in E_{_2}$, there exists an edge $e_1=(u_1,w_2)\in E_{_1}$ with: $(w_1,w_2)\in S$, $\mathcal{L}_{_1}(e_1)=\mathcal{L}_{_2}(e_2)$, and $\mathcal{C}_{_1}(e_1)\geq \mathcal{C}_{_2}(e_2)$.
 
 \item For each node $u_2\in V_{_2}$, there exists at least one node $u_1\in V_{_1}$ with $(u_1,u_2)\in S$.\hfill$\Square$
\end{enumerate}
\end{definition}

Intuitively, $Q_1\vartriangleright Q_2$ if there exists a subpattern of $Q_1$ that matches all constraints of $Q_2$ (i.e. labeling and attributes constraints, \emph{CQs} of edges, child and parent relationships). Thus, for any data graph $G$, if $G$ matches $Q_1$ then it matches $Q_2$ too. Moreover, if $G$ matches $Q_2$ then the corresponding match result may be refined to find that of $Q_1$ over $G$.

\begin{lemma}\label{lemma:Complexity_of_QPoM}
For any two QGPs $Q_1$ and $Q_2$, it is in $O(|Q_1|.|Q_2|)$ time to determine whether $Q_1\vartriangleright Q_2$ and if so, to compute the maximum match relation $S$. Moreover, there exists a unique maximum match relation $S$ for $Q_1$ in $Q_2$.\hfill$\Square$
\end{lemma}

We emphasize that the notion of \emph{pattern-only matching} is not equivalent to the notion of \emph{graph similarity} \cite{Henzinger,Adding_Reg_Exp_Fan_11} but it is inspired by it. We show later that our notion plays an important role when deciding containment of \emph{CGPs}.

\section{Traditional Containment of \emph{CGPs}}\label{section:tc}
In this section, we define the containment of \emph{CGPs} under the well-known traditional semantics, we discuss its checking time and we reveal its limit.

\subsection{Definition \& Checking}
The containment problem is traditionally defined for different kind of queries (e.g. relational queries \cite{Relational_Containment_1}, XPath queries \cite{XPath_Containment_2}, reachability queries \cite{Containment_of_RGQs}) as follows. A query $Q_1$ is \emph{contained} in a query $Q_2$ if, for any data instance $D$, $Q_1(D)\subseteq Q_2(D)$ where $Q_{i\in[1,2]}(D)$ is the result of evaluating $Q_i$ on $D$. When considering data and patterns modeled as graphs, evaluating a graph pattern $Q_2$ over a data graph $G$ by simulation \cite{GraphSimulation} and all its variants \cite{Fan14, Adding_Reg_Exp_Fan_11, Mahfoud18,Mahfoud20,MahfoudMedPrai20} yields a match result defined as a function that maps each node (resp. edge) of $Q_2$ to all its matches in $G$. Therefore, we revise the traditional definition of containment for \emph{CGPs} in terms of match result as follows.

\begin{definition}\label{definition:traditional_containment_of_cgps}
For any two CGPs $C_{i\in[1,2]}=(V_{_i},E_{_i},\mathcal{L}_{_i},\mathcal{A}_{_i},\mathcal{C}_{_i},\mathcal{P}^{+}_{_i},\mathcal{P}^{-}_{_i})$, we say that $C_{_1}$ is \emph{contained} in $C_{_2}$ via conditional simulation, denoted by $C_{_1}\sqsubseteq C_{_2}$, if there exists a mapping $\lambda$ from the core nodes (resp. core edges) of $C_{_1}$ to the core nodes (resp. core edges) of $C_{_2}$ such that: for any data graph $G$ and any node $u\in V_{_1}$ (resp. edge $e\in E_{_1}$), $\mathcal{M}^{G}_{_{C_{_1}}}(u)\subseteq \mathcal{M}^{G}_{_{C_{_2}}}(\lambda(u))$ (resp. $\mathcal{M}^{G}_{_{C_{_1}}}(e)\subseteq \mathcal{M}^{G}_{_{C_{_2}}}(\lambda(e))$).\footnote{Notice that if $\lambda(u)=\{u_1,\cdots,u_n\}$ then 
$\mathcal{M}^{G}_{_{C_{_2}}}(\lambda(u))=\mathcal{M}^{G}_{_{C_{_2}}}(u_1)\cup\cdots\cup \mathcal{M}^{G}_{_{C_{_2}}}(u_n)$, and similarly for any edge mapping $\lambda(e)=\{e_1,\cdots,e_n\}$.}\hfill$\Square$
\end{definition}


Recall that when evaluating a \emph{CGP} $C_{_2}$ over a data graph $G$, the match result $\mathcal{M}^{G}_{_{C_{_2}}}$ contains only matches of nodes and edges that belong to the core of $C_{_2}$ (i.e. the sets $V_{_2}$ and $E_{_2}$), while matches of any predicate are not present in $\mathcal{M}^{G}_{_{C_{_2}}}$. That is why we say that $C_{_1}$ is contained in $C_{_2}$ only if each core node (resp. core edge) in $C_{_1}$ is mapped to at least one core node (resp. core edge) in $C_{_2}$, in this way, matches of $C_{_1}$ over any data graph $G$ are all returned by $C_{_2}$ over $G$.

\begin{example}\label{example:traditional_containment_of_cgps}
Consider the patterns $C_{i\in[1,5]}$ of Example \ref{example:introduction}. It is easy to see that there exists a mapping $\lambda$ between core nodes (resp. edges) of $C_1$ and core nodes (resp. edges) of $C_2$ such that: $\lambda(a_1)=a_2$, $\lambda(b_1)=b_2$ and $\lambda(a_1\rightarrow b_1)=a_2\rightarrow b_2$. In other words, for any data graph $G$, $\mathcal{M}^{G}_{_{C_{_1}}}(a_1)\subseteq \mathcal{M}^{G}_{_{C_{_2}}}(a_2)$, $\mathcal{M}^{G}_{_{C_{_1}}}(b_1)\subseteq \mathcal{M}^{G}_{_{C_{_2}}}(b_2)$ and $\mathcal{M}^{G}_{_{C_{_1}}}(a_1\rightarrow b_1)\subseteq \mathcal{M}^{G}_{_{C_{_2}}}(a_2\rightarrow b_2)$. Thus, we conclude that $C_{_1}\sqsubseteq C_{_2}$. 
With the same principle, one can check that: $C_{_3}\sqsubseteq C_{_4}$, $C_{_5}\sqsubseteq C_{i\in[1,3,4]}$ but $C_{_4}\not\sqsubseteq C_{_3}$ (the edge $b_4\rightarrow c_4$ cannot be mapped to any core edge in $C_{3}$).\hfill$\Square$
\end{example}

We next give necessary and sufficient conditions to check the traditional containment of \emph{CGPs}.

\begin{lemma}\label{lemma:traditional_containment_of_cgps_nsc}
For any two CGPs $C_{i\in[1,2]}=(V_{_i},E_{_i},\mathcal{L}_{_i},\mathcal{A}_{_i},\mathcal{C}_{_i},\mathcal{P}^{+}_{_i},\mathcal{P}^{-}_{_i})$, $C_{_1}\sqsubseteq C_{_2}$ if and only if there exists a match relation $S\subseteq V_{_1}\cup V^{+}_{_1}\times V_{_2}\cup V^{+}_{_2}$ such that:

\begin{enumerate}
 \item $C^{+}_{_1}\vartriangleright C^{+}_{_2}$ with $S$;
 
 \item for each $(u_1,u_2)\!\in\! S$ and each negative predicate $p^{-}_2(u_2)\!\in\!\mathcal{P}^{-}_{_2}$, there exists a negative predicate $p^{-}_{1}(u_1)\!\in\!\mathcal{P}^{-}_{_1}$ where: $p^{-}_{2}\vartriangleright p^{-}_{1}$ with some match relation $S^{'}$ and $(u_1,u_2)\in S^{'}$;
 
  \item for any core node $u_1\in V_{_1}$, there exists at least one core node $u_2\in V_{_2}$ with $(u_1,u_2)\in S$. Moreover, for any core edge $(u_1,w_1)\in E_{_1}$, there exists at least one core edge $(u_2,w_2)\in E_{_2}$ with $\{(u_1,u_2),(w_1,w_2)\}\in S$.
\end{enumerate}
We say that the match relation $S$ realizes $\lambda$.\hfill$\Square$
\end{lemma}

When $C_{_1}\sqsubseteq C_{_2}$ for two \emph{CGPs} $C_{_1}$ and $C_{_2}$, the match result of $C_{_2}$ over any data graph $G$ can be considered as an over-evaluation of $C_{_1}$ on $G$. Rather to return $\mathcal{M}^{G}_{C_{_2}}$ naively, some optimization can be done over it as follows.

\begin{proposition}\label{proposition:reduce_unnecessary_matches}
\sloppy{For any data graph $G$ and any CGPs $C_{i\in[1,2]}=(V_{_i},E_{_i},\mathcal{L}_{_i},\mathcal{A}_{_i},\mathcal{C}_{_i},\mathcal{P}^{+}_{_i},\mathcal{P}^{-}_{_i})$ where $C_{_1}\sqsubseteq C_{_2}$ via the mapping $\lambda$, we have: $v\in\mathcal{M}^{G}_{C_{_1}}(u)$, for some data graph $v$ and a core node $u$ in $C_{_1}$, if and only if $v\in\bigcap_{u'\in\lambda(u)}\mathcal{M}^{G}_{C_{_2}}(u')$. Moreover, $e_v\in\mathcal{M}^{G}_{C_{_1}}(e)$, for some data edge $e_v$ and a core edge $e$ in $C_{_1}$, if and only if $e_v\in\bigcap_{e'\in\lambda(e)}\mathcal{M}^{G}_{C_{_2}}(e)$.}\hfill$\Square$
\end{proposition}


Finally, the checking time of the traditional containment is stated as follows.

\begin{theorem}\label{theorem:complexity_traditional_containment_CGPs}
For any two CGPs $C_{i\in[1,2]}$, it is in $O(|C_{_1}|.|C_{_2}|)$ time to check whether $C_{_1}\sqsubseteq C_{_2}$ and if so, to compute the corresponding mapping $\lambda$.\hfill$\Square$
\end{theorem}

We prove Theorem \ref{theorem:complexity_traditional_containment_CGPs} by providing a quadratic-time algorithm to check traditional containment between two \emph{CGPs}. Our algorithm, referred to as \textsf{TContained}, is shown in Fig. \ref{Algorithm_TContained}. Given two \emph{CGPs} $C_{_1}$ and $C_{_2}$ in input, it returns $\emptyset$ if $C_{_1}\not\sqsubseteq C_{_2}$, or the mapping $\lambda$ that allows $C_{_1}$ to be traditionally contained in $C_{_2}$ as well as the match relation $S$ that realizes $\lambda$. Description and complexity analysis of \textsf{TContained} are given in 
Appendix.

\begin{figure}[t!]
\rule{\linewidth}{0.5pt}
\scriptsize
\begin{flushleft}
\textbf{Algorithm} \textsf{TContained}($C_{_1},C_{_2}$)\newline
\textit{Input}: Two \emph{CGPs} $C_{i\in[1,2]}$=($V_{_i},E_{_i},\mathcal{L}_{_i},\mathcal{A}_{_i},\mathcal{C}_{_i},\mathcal{P}^{+}_{_i},\mathcal{P}^{-}_{_i}$).\newline
\textit{Output}: A pair ($\lambda, S$) if $C_{_1}\sqsubseteq C_{_2}$ via $\lambda$ and $S$ is the match relation $S$ that realizes $\lambda$; or $\emptyset$ otherwise.

\end{flushleft}
\begin{algorithmic}[1]
\State $S$ := \textsf{POM}($C^{+}_{_1}$, $C^{+}_{_2}$);
\ForAll{$(u_1,u_2)\in S$ \textbf{and each} $p^{-}_2(u_2)\in\mathcal{P}^{-}_{_2}$}
    \If{($\nexists p_1^{-}(u_1)\in\mathcal{P}^{-}_{_1}$ \textbf{s.t:} $S'$=\textsf{POM}($p_2^{-}(u_2),p_1^{-}(u_1)$) \textbf{$\&$} $(u_1,u_2)\in S'$)}
	  \State $S$ := $S\setminus \{(u_1,u_2)\}$;
      \EndIf
\EndFor

\ForAll{$(u_1,u_2)\in S$ \textbf{s.t:} $u_1\in V_ {_1}$ \textbf{$\&$} $u_2\in V^{+}_{_2}$}
    \State $S$ := $S\setminus \{(u_1,u_2)\}$;
\EndFor

\State $S$ := \textsf{POM}($C^{+}_{_1}$, $C^{+}_{_2}, S$);
\If{$S=\emptyset$}
    \Return ($\emptyset,\emptyset$);
\EndIf

\ForAll{$u_1\in V_{_1}$}
    \State $\lambda(u_1) := \emptyset$;
    \ForAll{$(u_1,u_2)\in S$ \textbf{with} $u_2\in V_{_2}$}
	  \State $\lambda(u_1) := \lambda(u_1)\cup \{u_2\}$;
    \EndFor
\EndFor

\ForAll{$e_1\in E_{_1}$ \textbf{with} $e_1=(u_1,w_1)$}
    \State $\lambda(e_1) := \emptyset$;
     \ForAll{$e_2\in E_{_2}$ \textbf{with} $e_2=(u_2,w_2)$ \textbf{$\&$} $\{(u_1,u_2),(w_1,w_2)\}\in S$}
	  \State $\lambda(e_1) := \lambda(e_1)\cup \{e_2\}$;
    \EndFor
\EndFor

\If{($\exists x\in V_{_1}$ (resp. $x\in E_{_1}$) \textbf{with} $\lambda(x)=\emptyset$)}
    \State \Return ($\emptyset,\emptyset$);
\Else
    \State \Return ($\lambda,S$);
\EndIf

\normalsize
\end{algorithmic}

\rule{\linewidth}{0.5pt}
\caption{Algorithm for traditional containment checking.}
\label{Algorithm_TContained}
\end{figure}
\nointerlineskip

\subsection{Equivalence}
It is well known that the equivalence between two patterns can be checked via a bidirectional containment. That is, for any two \emph{CGPs} $C_1$ and $C_2$, $C_1$ is equivalent to $C_2$, denoted by $C_1\equiv C_2$, iff: $C_1\sqsubseteq C_2$ and $C_2\sqsubseteq C_1$. When it comes to \emph{QGPs} with query focus, we slightly revise the semantics of equivalence as follows.

\begin{definition}\label{definition:equivalence_of_QGPs}
Given two QGPs $Q_1(u_1)$ and $Q_2(u_2)$ where $Q_{i\in[1,2]}=(V_{_i},E_{_i},\mathcal{L}_{_i},\mathcal{A}_{_i},\mathcal{C}_{_i})$, $u_1\in V_{_1}$ and $u_2\in V_{_2}$. We say that $Q_1(u_1)$ and $Q_2(u_2)$ are \emph{equivalent}, written $Q_1(u_1)\equiv Q_2(u_2)$, if:

\begin{enumerate}
 \item $Q^{'}_1\sqsubseteq Q^{'}_2$ with the maximum match relation $S_{1\rightarrow 2}$;
 \item $Q^{'}_2\sqsubseteq Q^{'}_1$ with the maximum match relation $S_{2\rightarrow 1}$; and
 \item $(u_1,u_2)\in S_{1\rightarrow 2}\bigwedge S_{2\rightarrow 1}$.
\end{enumerate}

\noindent Where $Q^{'}_{i\in[1,2]}$ is obtained by replacing $\mathcal{A}_{i}(u_{i})$ by $\emptyset$ in $Q_{i}$.\hfill$\Square$
\end{definition}

Informally, the equivalence between $Q_1(u_1)$ and $Q_2(u_2)$ is checked by ignoring the attributes constraints defined over the query focus $u_1$ and $u_2$. 

\begin{figure}[t!]
\centering
   \noindent\makebox[\textwidth]{%
   \includegraphics[width=1.6\linewidth,height=9.2cm]{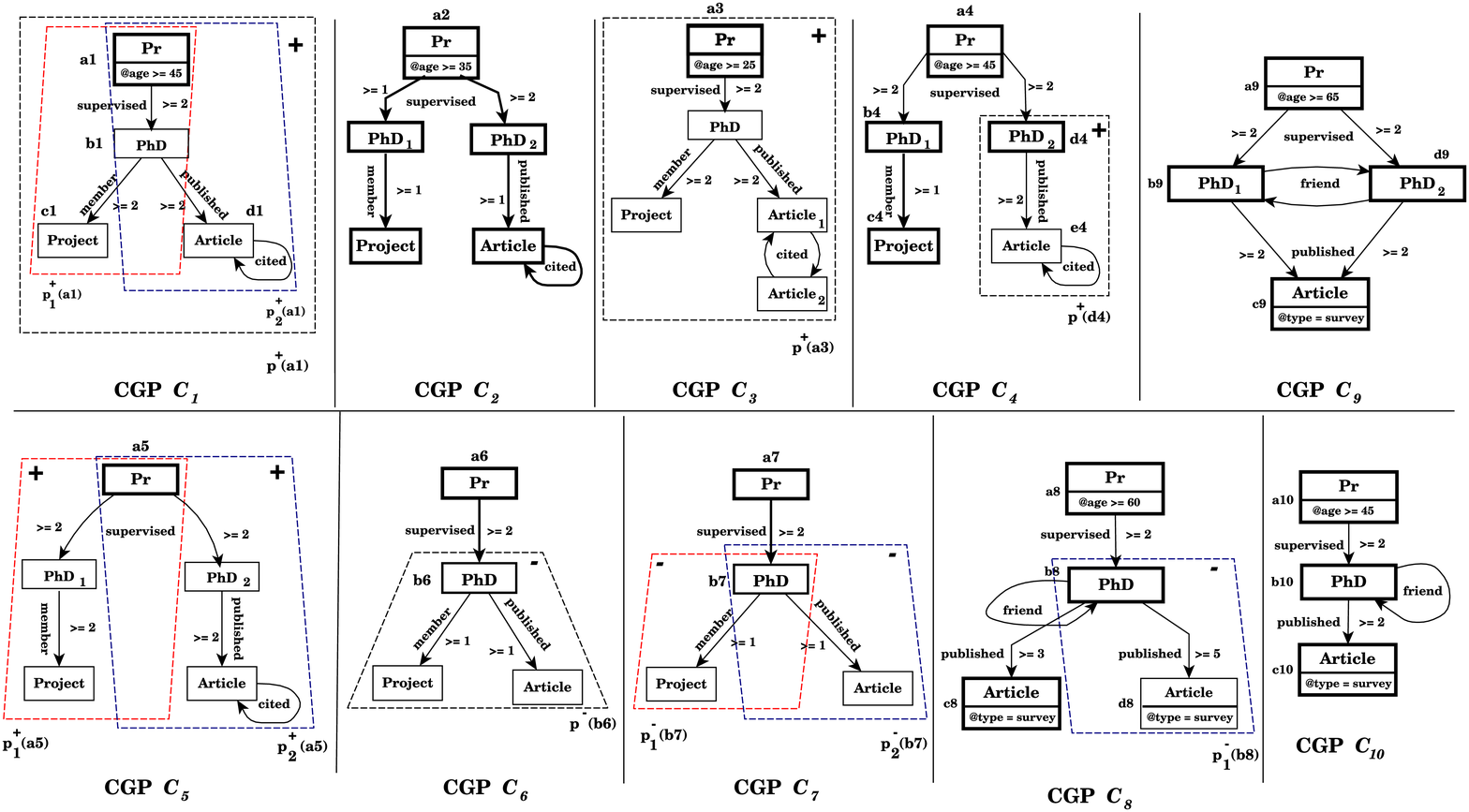}}
   \caption{Example of \emph{CGPs}.}
   \label{figure:example_of_cgps}
\end{figure}

\begin{example}
Consider the \emph{QGPs} $C_9$ and $C_{10}$ of Fig. \ref{figure:example_of_cgps}. Remark that $C_9\not\equiv C_{10}$ since the sets of professors looked up by $C_9$ and $C_{10}$ are not the same. By considering query focus, one can check that $C_9(a_9)\equiv C_{10}(a_{10})$ since, when ignoring the attributes constraints defined over the nodes $a_9$ and $a_{10}$, the remaining parts of $C_9$ and $C_{10}$ are equivalent. However, $C_9(c_9)\not\equiv C_{10}(c_{10})$.
\hfill$\Square$
\end{example}

\subsection{Limit of Traditional Containment}
The traditional definition of containment applies well for patterns without predicates and allows to exactly match some pattern based only on the match result of another one. 
For patterns with predicates however, the traditional containment may allow to do only \emph{sub-matching/over-matching} but not necessarily exact matching. We show this limit by the next example.


\begin{example}\label{example:limit_of_traditional_containment}
We have shown in Example \ref{example:traditional_containment_of_cgps} that, for any data graph $G$, $\mathcal{M}^{G}_{_{C_{1}}}(a_1)\subseteq \mathcal{M}^{G}_{_{C_{_2}}}(a_2)$ and $\mathcal{M}^{G}_{_{C_{_1}}}(b_1)\subseteq \mathcal{M}^{G}_{_{C_{_2}}}(b_2)$. The limit of this traditional containment is that the exact matches $\mathcal{M}^{G}_{_{C_{_1}}}(b_1)$ cannot be extracted from $\mathcal{M}^{G}_{_{C_{_2}}}(b_2)$ since: for each PhD student returned by $C_2$ (i.e. a match $v$ of the node $b_2$ in $\mathcal{M}^{G}_{_{C_{_2}}}(b_2)$), we must check whether he has at least two published articles, however, no information about published articles is returned by $C_2$ which makes impossible the evaluation of the positive predicate of $C_1$ over $v$. Therefore, even if $C_{_1}\sqsubseteq C_{_2}$, $\mathcal{M}^{G}_{_{C_{_1}}}$ cannot be extracted from $\mathcal{M}^{G}_{_{C_{_2}}}$, and moreover, the match result $\mathcal{M}^{G}_{_{C_{_2}}}$ can be considered as an \emph{over-matching} of $C_{_1}$ on $G$. Consider $C_3$ and $C_4$, and remark that: a) $\mathcal{M}^{G}_{_{C_{_3}}}(b_3)\subseteq \mathcal{M}^{G}_{_{C_{_4}}}(b_4)$; and b) each match $v$ of the node $b_4$ in $C_4$ is returned in $\mathcal{M}^{G}_{_{C_{_4}}}$ with its corresponding projects children which makes possible the evaluation of the predicate of $C_3$ over $v$. This means that, the exact match result $\mathcal{M}^{G}_{_{C_{_3}}}$ can be extracted from $\mathcal{M}^{G}_{_{C_{_4}}}$ over any data graph $G$.\hfill$\Square$
\end{example}

From this limit, a question arises: ``\emph{given two CGPs $C_{_1}$ and $C_{_2}$, in which cases the match result of $C_{_1}$ over any data graph $G$ can be extracted from that of $C_{_2}$ over $G$ ?}''. Answering this question will allow to optimize matching of \emph{CGPs} in emerging applications that require exact matching rather than sub/over matching. We tackle this problem in the next section.

\section{Revisited Containment for \emph{CGPs}}\label{section:sc}
The main result of this section is a new semantics for containment, called \emph{strong containment}, that overcomes the limit of traditional containment.

\subsection{Predicates Evaluability}
Given two \emph{CGPs} $C_{i\in[1,2]}$ such that $C_{_1}\sqsubseteq C_{_2}$ via the mapping $\lambda$. We have explained that, even $\mathcal{M}^{G}_{C_{_1}}(u)\subseteq \mathcal{M}^{G}_{C_{_2}}(\lambda(u))$ for any core node $u$ in $C_{_1}$ and any data graph $G$, it is not always possible to extract all matches of $u$ from $\mathcal{M}^{G}_{C_{_2}}(\lambda(u))$ in presence of predicates. To decide statically whether such extraction is possible, we introduce the notion of \emph{predicate evaluability}.

\begin{definition}\label{definition:evaluability_of_predicates}
Given two CGPs $C_{i\in[1,2]}=(V_{_i},E_{_i},\mathcal{L}_{_i},\mathcal{A}_{_i},\mathcal{C}_{_i},\mathcal{P}^{+}_{_i},\mathcal{P}^{-}_{_i})$ where $C_{_1}\sqsubseteq C_{_2}$ via the mapping $\lambda$, and consider a node $u\in V_{1}$ where $\lambda(u)=\{u_1,\cdots,u_n\}$. A predicate $p(u)\in\mathcal{P}^{+}_{_1}\cup \mathcal{P}^{-}_{_1}$ (i.e. positive or negative) is \emph{evaluable} over $\lambda(u)$ if, for any data graph $G$ and any data node $v\in\bigcap_{u'\in\lambda(u)}\mathcal{M}^{G}_{C_{_2}}(u')$, it can be decided whether $v$ satisfies $p(u)$.\hfill$\Square$
\end{definition}

The intersection is due to the result of Proposition \ref{proposition:reduce_unnecessary_matches}. Obviously, the \emph{evaluability} aims to check whether the potential match set of $u$, represented by $\mathcal{M}^{G}_{C_{_2}}(u_1)\cap\cdots\cap \mathcal{M}^{G}_{C_{_2}}(u_n)$, can be refined to keep only data nodes that satisfy the predicate $p(u)$. 

The semantics of \emph{predicates evaluability} is quite simple, but its implementation is not trivial and requires to introduce two more notions, \emph{refinement} and \emph{elimination}, that we explain by the next example.

\begin{example}\label{example:predicate_evaluability}
\noindent\underline{\emph{Case of positive predicates.}} Consider the \emph{CGPs} $C_{i\in[1,5]}$ of Fig. \ref{figure:example_of_cgps} and remark that $C_{1}\sqsubseteq C_i$ for any $i\in[2,5]$. For any data graph $G$, it is easy to see that the matches of the node $Pr$ in $\mathcal{M}^{G}_{C_{2}}$ can be refined to keep only those that satisfy the predicate $p^{+}(a_1)$ of $C_{1}$. Thus, we say that $p^{+}(a_1)$ can be evaluated by \emph{refinement} over the match result $\mathcal{M}^{G}_{C_{_2}}$. Remark that $p^{+}(a_3)\equiv p^{+}(a_1)$, which means that for any match $v$ of the node \emph{Pr} in $\mathcal{M}^{G}_{C_{_3}}$, $v$ is a match of $C_{_1}$ only if it satisfies the attribute ``\emph{@age$>45$}'' of $C_{1}$. Then, matches of $C_{1}$ can be extracted by refining $\mathcal{M}^{G}_{C_{_3}}$ w.r.t the attribute ``\emph{@age$\geq 45$}'' while the predicate $p^{+}(a_1)$ can be \emph{eliminated} from this process. These \emph{refinement} and \emph{elimination} notions are combined together in the case of $C_{4}$. $C_{4}$ contains two nodes with the same label \emph{PhD}, the first one ($PhD_1$) must have at least one \emph{Project} child, while the second one ($PhD_2$) must satisfy the predicate $p^{+}(d_4)$. It is clear that $p^{+}(d_4)$ is equivalent to some part of $p^{+}(a_1)$, in other words, this part is eliminable over matches of $PhD_2$. Thus, matches of $p^{+}(a_1)$ can be obtained by refining matches of $PhD_1$ and by combining them together with the matches of $PhD_2$ (without refining these later). We say that $p^{+}(a_1)$ is evaluated over $C_{4}$ both by \emph{refinement \& elimination}.
\noindent\underline{\emph{Case of negative predicates.}} Consider now the \emph{CGPs} $C_{6}$ and $C_{7}$ of Fig. \ref{figure:example_of_cgps} and remark that the predicate $p^{-}(b_6)$ in $C_{6}$ is splitted into two negative predicates, $p^{-}_{1}(b_7)$ and $p^{-}_{2}(b_7)$, in $C_{_7}$. Contrary to positive predicates, even the part $\scriptstyle{PhD\xrightarrow[member]{\geq 1}Project}$ (resp. $\scriptstyle{PhD\xrightarrow[published]{\geq 1}Article}$) of $p^{-}(b_6)$ is equivalent to $p^{-}_{1}(b_7)$ (resp. $p^{-}_{2}(b_7)$), we cannot eliminate the predicate $p^{-}(b_6)$ over matches of $PhD$ in $\mathcal{M}^{G}_{C_{_7}}$ since there may be some data nodes that satisfy $p^{-}(b_6)$ but which are not returned in $\mathcal{M}^{G}_{C_{_7}}$. The reason why we cannot eliminate a negative predicate over a conjunction of its parts, like for positive predicates, is due to the fact that this conjunction changes completely its semantics.\hfill$\Square$
\end{example}


We give next formal definition of the notions of \emph{refinement} and \emph{elimination}.

\begin{definition}\label{definition:refinement_elimination}
Given two CGPs $C_{i\in[1,2]}=(V_{_i},E_{_i},\mathcal{L}_{_i},\mathcal{A}_{_i},\mathcal{C}_{_i},\mathcal{P}^{+}_{_i},\mathcal{P}^{-}_{_i})$ where $C_{_1}\sqsubseteq C_{_2}$ via the mapping $\lambda$, and let $S$ be the match relation that realizes $\lambda$. We define a \emph{refinement} relation $\mathcal{R}^{+}$ (resp. $\mathcal{R}^{-}$) and an \emph{elimination} relation $\mathcal{E}^{+}$ (resp. $\mathcal{E}^{-}$) for positive (resp. negative) predicates as follows:

\begin{enumerate}
 \item for each $(e_1,e_2)\in \mathcal{R}^{+}$ with $e_1\!=\!(u_1,w_1)$ and $e_2\!=\!(u_2,w_2)$, we have: $e_1\in E^{+}_{_1}$; $e_2\in E_{_2}$ and $\{(u_1,u_2),(w_1,w_2)\}\in S$.
 
 \item for each $(p^{+}_1(u_1),u_2)\in\mathcal{E}^{+}$ we have:    
 \begin{enumerate}
  \item $p^{+}_1(u_1)$ is (a part of) a positive predicate in $\mathcal{P}^{+}_{_1}$; $(u_1,u_2)\in S$; $u_2\in V_{_2}$; and there exists (a part of) a positive predicate $p^{+}_2(u_2)$ in $\mathcal{P}^{+}_{_2}$ s.t: $p^{+}_1(u_1)\equiv p^{+}_2(u_2)$.
  
  \item for each predicate edge $e'_{1}\in E^{+}_{_1}$ that is not in $p^{+}_1(u_1)$, either $e'_{1}$ intersects with $p^{+}_1(u_1)$ only on $u_1$ or it does not intersect with $p^{+}_1(u_1)$.
 \end{enumerate}
 
  
  \item for each $(e,e')\in \mathcal{R}^{-}$ we have: a) $e\in E^{-}_{_1}$; b) $e'\in E_{_2}$; c) $e\sim e'$; and d) there exists $(u,w)\in S$; an undirected paths $\textlangle e_1,\cdots,e_n,e\textrangle$ in $p^{-}_{1}(u_1)$ that starts at $u$; an undirected path $\textlangle e'_1,\cdots,e'_n,e'\textrangle$ in $C_{_2}$ that starts at $w$ where: $(e_i,e'_i)\in \mathcal{R}^{-}$ for each $1\leq i\leq n$.
  

 \item for each $(p^{-}_{1}(u_1),u_2)\in\mathcal{E}^{-}$ we have: $p^{-}_{1}(u_1)\in\mathcal{P}^{-}_{_1}$; $(u_1,u_2)\in S$; and there exists $p^{-}_2(u_2)\in\mathcal{P}^{-}_{_2}$ s.t: $p^{-}_{1}(u_1)\equiv p^{-}_2(u_2)$.\hfill$\Square$
\end{enumerate}
\end{definition}

Intuitively, the relation $\mathcal{R}^{+}$ (resp. $\mathcal{R}^{-}$) contains all predicates edges in $E^{+}_{_1}$ (resp. $E^{-}_{_1}$) that are mapped to core edges in $C_{_2}$. This means that matches of these negative edges can be found by \emph{refining} matches of some core edges in $C_{_2}$. Given a (part of) positive predicate $p^{+}_1(u_1)$ in $C_{_1}$, the relation $\mathcal{E}^{+}$ maps $p^{+}_1(u_1)$ to a core node $u_2$ from $C_{_2}$ if there exists a (part of) positive predicate $p^{+}_2(u_2)$ in $C_{_2}$ that is equivalent to $p^{+}_1(u_1)$. In this way, when refining matches of $u_2$ to find those of $u_1$, $p^{+}_1(u_1)$ can be \emph{eliminated} from this \emph{refinement} since each match of $u_2$ satisfies it. The same principle is applied for negative predicates with the relation $\mathcal{E}^{-}$. We emphasize that, contrary to positive predicates, a negative predicate may be completely \emph{refinable} via the relation $\mathcal{R}^{-}$ or  completely \emph{eliminable} via the relation $\mathcal{E}^{-}$ while partial elimination/refinement can change its semantics (see Example \ref{example:predicate_evaluability}). Moreover, when eliminating parts of a positive predicate in $C_{_1}$ over core nodes in $C_{_2}$, the connectivity defined between these parts in $C_{_1}$ can be lost, hence, condition (2-b) allows to preserve this connectivity (see next example).

\begin{example}\label{example:formal_refinement_elimination}
Consider the patterns of Fig. \ref{figure:example_of_cgps}. 
We denote by $p^{+}_{3}(a_1)$ the part of the predicate $p^{+}(a_1)$ in $C_1$ that is composed by the nodes $b_1$ and $d_1$ with all edges connecting them. We have: ${\scriptstyle \mathcal{R}^{+}_{1\rightarrow 4}=\{(a_1\rightarrow b_1,a_4\rightarrow b_4),(a_1\rightarrow b_1,a_4\rightarrow d_4),(b_1\rightarrow c_1,b_4\rightarrow c_4)\}}$; ${\scriptstyle \mathcal{E}^{+}_{1\rightarrow 4}=\{(p^{+}_{3}(a_1),d_4)\}}$ since the positive predicate $p^{+}(d_4)$ defined over the node $d_4$ in $C_{4}$ is equivalent to the part $p^{+}_{3}(a_1)$ of $p^{+}(a_1)$; and ${\scriptstyle \mathcal{E}^{-}_{1\rightarrow 4} = \mathcal{R}^{-}_{1\rightarrow 4}} = \emptyset$. Hence, the part $p^{+}_{3}(a_1)$ of $p^{+}(a_1)$ in $C_{1}$ is eliminable over matches of the node $d_4$ of $C_{4}$, while the remaining parts of $p^{+}(a_1)$ are refinable over matches of $C_{4}$. By considering $C_{1}$ over $C_{5}$, we have: ${\scriptstyle \mathcal{R}_{1\rightarrow 5}=\mathcal{E}^{+}_{1\rightarrow 5} = \mathcal{E}^{-}_{1\rightarrow 5} = \mathcal{R}^{-}_{1\rightarrow 5}} = \emptyset$. Remark that the part $p^{+}_1(a_1)$ (resp. $p^{+}_2(a_1)$) of the predicate $p^{+}(a_1)$ in $C_1$ is equivalent to the part $p^{+}_1(a_5)$ (resp. $p^{+}_2(a_5)$) of the predicate $p^{+}(a_5)$ in $C_{_5}$ but ${\scriptstyle \{(p^{+}_1(a_1), a_5),(p^{+}_2(a_1), a_5)\}\not\subseteq\mathcal{E}^{+}_{1\rightarrow 5}}$ since matches of the predicates nodes $PhD_1$ and $PhD_2$ in $C_{5}$ are not returned within the match result of $C_{5}$, thus we cannot combine their matches to find those of the node $PhD$ in $C_{1}$. In other words, the connectivity between the parts $p^{+}_1(a_1)$ and $p^{+}_2(a_1)$ of $p^{+}(a_1)$ will be lost if we eliminate them over the predicates $p^{+}_1(a_5)$ and $p^{+}_1(a_5)$ of $C_{_5}$. That is why, condition (2b) of Def. \ref{definition:refinement_elimination} allows to eliminate parts of a positive predicate only if a combination can be done later to find matches of the whole predicate. For $C_{8}$ and $C_{10}$, we have: ${\scriptstyle \mathcal{R}^{+}_{8\rightarrow 10}=\{(a_8\rightarrow b_8,a_{10}\rightarrow b_{10}),(b_8\rightarrow c_{10},b_{10}\rightarrow c_{10})\}}$; ${\scriptstyle \mathcal{R}^{-}_{8\rightarrow 10}=\{(b_8\rightarrow d_8,b_{10}\rightarrow c_{10})\}}$; ${\scriptstyle \mathcal{E}^{+}_{8\rightarrow 10} = \mathcal{E}^{-}_{8\rightarrow 10}} = \emptyset$. Thus, matches of the negative predicate $p^{-}(b_8)$ of $C_{_8}$ can be found by refining matches of the edge $b_{10}\rightarrow c_{10}$ of $C_{_{10}}$. \hfill$\Square$
\end{example}

We give next necessary and sufficient conditions for predicates evaluability.

\begin{lemma}\label{lemma:ns_conditions_evaluability_predicates}
Given two CGPs $C_{i\in[1,2]}=(V_{_i},E_{_i},\mathcal{L}_{_i},\mathcal{A}_{_i},\mathcal{C}_{_i},\mathcal{P}^{+}_{_i},\mathcal{P}^{-}_{_i})$ where $C_{_1}\sqsubseteq C_{_2}$ via the mapping $\lambda$. Let $\mathcal{R}^{+}$, $\mathcal{R}^{-}$, $\mathcal{E}^{+}$ and $\mathcal{E}^{-}$ be the corresponding refinement and elimination relations as defined in Def. \ref{definition:refinement_elimination}. A positive predicate $p^{+}(u)$ is \emph{evaluable} over $\lambda(u)$ if each edge in $p^{+}(u)$ is covered by $\mathcal{R}^{+}$ or $\mathcal{E}^{+}$. Moreover, a negative predicate $p^{-}_{1}(u)$ is \emph{evaluable} over $\lambda(u)$ if: a) there exists a pair $(p^{-}_{1}(u),u')\in\mathcal{E}^{-}$ for some node $u'\in\lambda(u)$; or b) all edges in $p^{-}_{1}(u)$ are covered by $\mathcal{R}^{-}$.\hfill$\Square$
\end{lemma}

\begin{example}\label{example:evaluability}
Consider the \emph{CGPs} $C_{i\in[1,10]}$ of Examples \ref{example:predicate_evaluability} and \ref{example:formal_refinement_elimination}. One can check that the predicate $p^{+}(a_1)$ in $C_{1}$ is \emph{evaluable}: a) by \emph{refinement} over the node $a_2$ of $C_2$; b) by \emph{elimination} over the node $a_3$ of $C_3$; and c) by \emph{refinement \& elimination} over $C_4$. Moreover, $p^{+}(a_1)$ is not \emph{evaluable} over the node $a_5$ of $C_5$. In addition, the predicate $p^{-}(b_8)$ in $C_{8}$ is \emph{evaluable} by \emph{refinement} over the node $b_{10}$ of $C_{10}$.\hfill$\Square$
\end{example}


\begin{figure}[t!]
\rule{\linewidth}{0.5pt}
\scriptsize
\begin{flushleft}
\textbf{Algorithm} \textsf{SContained}($C_{_1},C_{_2}$)\newline
\textit{Input}: Two \emph{CGPs} $C_{i\in[1,2]}$=($V_{_i},E_{_i},\mathcal{L}_{_i},\mathcal{A}_{_i},\mathcal{C}_{_i},\mathcal{P}^{+}_{_i},\mathcal{P}^{-}_{_i}$).\newline
\textit{Output}: ($\lambda, \mathcal{R},\mathcal{E}^{+},\mathcal{E}^{-}$) if $C_{_1}\sqsubseteq_s C_{_2}$; or $\emptyset$ otherwise.
\end{flushleft}
\begin{algorithmic}[1]
\State ($\lambda,S$) := \textsf{TContained}($C_{_1}$, $C_{_2}$);
\If{(($\lambda, S$) $=$ ($\emptyset,\emptyset$))}
     \Return $\emptyset$;
\EndIf


\State ($\mathcal{R}^{+},\mathcal{E}^{+},\mathcal{R}^{-},\mathcal{E}^{-}$) := \textsf{ExtractRels}($C_{_1},C_{_2},S$);

\State $CE^{+} := \emptyset$, $CE^{-} := \emptyset$;

\ForAll{$(e_1,e_2)\in \mathcal{R}^{+}$}
    $CE^{+} := CE^{+} \cup \{e_1\}$;
\EndFor

\ForAll{$(p^{+}_1(u_1),u_2)\in \mathcal{E}^{+}$ \textbf{and each} edge $e\in p^{+}_1(u_1)$}
    $CE^{+} := CE^{+} \cup \{e\}$;
\EndFor

\ForAll{$(e_1,e_2)\in \mathcal{R}^{-}$}
    $CE^{-} := CE^{-} \cup \{e_1\}$;
\EndFor

\ForAll{$(p^{-}_{1}(u_1),u_2)\in \mathcal{E}^{-}$ \textbf{and each} edge $e\in p^{-}_{1}(u_1)$}
    $CE^{-} := CE^{-} \cup \{e\}$;
\EndFor

\If{($CE^{+}=E^{+}_{_1}$ \textbf{and} $CE^{-}=E^{-}_{_1}$)}
    \State \Return ($\lambda, \mathcal{R}^{+},\mathcal{R}^{-}$);
\Else
    \State \Return $\emptyset$;
\EndIf

\normalsize
\end{algorithmic}

\rule{\linewidth}{0.5pt}
\caption{Algorithm for \emph{strong containment} checking.}
\label{Algorithm_SContained}
\end{figure}
\nointerlineskip

\subsection{Strong Containment}
Based on the notion of \emph{predicates evaluability}, we give a revisited semantics of traditional containment as follows.

\begin{definition}\label{definition:strong_containment_of_cgps}
For any CGPs $C_{i\in[1,2]}=(V_{_i},E_{_i},\mathcal{L}_{_i},\mathcal{A}_{_i},\mathcal{C}_{_i},\mathcal{P}^{+}_{_i},\mathcal{P}^{-}_{_i})$, $C_{_1}$ is \emph{strongly contained} in $C_{_2}$ via conditional simulation, written $C_{_1}\sqsubseteq_{s} C_{_2}$, if: a) $C_{_1}\sqsubseteq C_{_2}$ via $\lambda$; and
b) each predicate $p(u)\in\mathcal{P}^{+}_{_1}\cup\mathcal{P}^{-}_{_1}$ is evaluable over $\lambda(u)$.\hfill$\Square$
\end{definition}

Indeed, \emph{strong containment} extends the traditional containment by condition (2) to check whether $\mathcal{M}^{G}_{_{C_{_1}}}$ can be extracted from $\mathcal{M}^{G}_{_{C_{_2}}}$ over any data graph $G$.

\begin{example}
Based on Examples \ref{example:predicate_evaluability}, \ref{example:formal_refinement_elimination} and \ref{example:evaluability}, one can check that: $C_{1}\sqsubseteq_s C_{i\in[2,4]}$, $C_{1}\not\sqsubseteq_s C{_5}$, $C_{7}\not\sqsubseteq_s C{_6}$ and $C_{8}\sqsubseteq_s C_{10}$. Thus, the match result $\mathcal{M}^{G}_{_{C_{_1}}}$ (resp. $\mathcal{M}^{G}_{_{C_{_8}}}$) can be extracted from $\mathcal{M}^{G}_{_{C_{i\in[2,4]}}}$ (resp. $\mathcal{M}^{G}_{_{C_{_{10}}}}$) over any data graph $G$.\hfill$\Square$
\end{example}

Necessary and sufficient conditions for checking strong containment follow from Lemmas \ref{lemma:traditional_containment_of_cgps_nsc} and \ref{lemma:ns_conditions_evaluability_predicates}. The main result of this section is stated as follows:

\begin{theorem}\label{theorem:complexity_strong_containment_CGPs}
For any two CGPs $C_{i\in[1,2]}$, it is in $O(|C_{_1}|.|C_{_2}|+|V^{+}_{_1}|.|\mathcal{P}^{+}_{_1}|.|\mathcal{P}^{+}_{_2}|)$ time to decide whether $C_{_1}\sqsubseteq_s C_{_2}$ and if so, to find the corresponding mapping $\lambda$ as well as the relations $\mathcal{R}$, $\mathcal{E}^{+}$ and $\mathcal{E}^{-}$.\hfill$\Square$
\end{theorem}


We prove Theorem \ref{theorem:complexity_strong_containment_CGPs} by providing an algorithm that checks strong containment of \emph{CGPs} in cubic-time. Our algorithm, referred to as \textsf{SContained}, is shown in Fig. \ref{Algorithm_SContained}. Given two \emph{CGPs} $C_{_1}$ and $C_{_2}$ in input, it invokes \textsf{TContained} to check whether $C_{_1}\sqsubseteq C_{_2}$ and to find the corresponding mapping $\lambda$ and the match relation $S$ that realizes it (line 1). If $C_{_1}\not\sqsubseteq C_{_2}$ then $\emptyset$ is returned (line 2), otherwise, the refinement and elimination relations (i.e. $\mathcal{R}^{+}$, $\mathcal{R}^{-}$, $\mathcal{E}^{+}$ and $\mathcal{E}^{-}$) are extracted by using the procedure \textsf{ExtractRels} (line 3) that is given in Appendix. The set $CE^{+}$ is defined (lines 5--6) to determine all positive edges in $E^{+}_{_1}$ that are covered by the relations $\mathcal{R}^{+}$ and/or $\mathcal{E}^{+}$. Similarly, the set $CE^{-}$ is defined (lines 7--8) to capture all negative edges in $E^{-}_{_1}$ that are covered by the relations $\mathcal{R}^{-}$ and/or $\mathcal{E}^{-}$. If each positive (resp. negative) edge in $\mathcal{P}^{+}_1$ (resp. $\mathcal{P}^{-}_1$) belongs to the set $CE^{+}$ (resp. $CE^{-}$) then all predicates edges in $C_{_1}$ are concerned by refinement and/or elimination, and thus, $C_{_1}\sqsubseteq_s C_{_2}$. In this case, \textsf{SContained} returns the mapping $\lambda$ and the refinement relations $\mathcal{R}^{+}$ and $\mathcal{R}^{-}$ (lines 9--10). Otherwise, it returns $\emptyset$ (lines 11--12). A detailed complexity analysis of algorithm \textsf{SContained} is given in Appendix to complete proof of Theorem \ref{theorem:complexity_strong_containment_CGPs}.

We show later that, when $C_{_1}\sqsubseteq_s C_{_2}$, only the mapping $\lambda$ and the refinement relations $\mathcal{R}^{+}$ and $\mathcal{R}^{-}$ are used to extract $\mathcal{M}^{G}_{_{C_{_1}}}$ from $\mathcal{M}^{G}_{_{C_{_2}}}$ over any data graph $G$, while the relations $\mathcal{E}^{+}$ and $\mathcal{E}^{-}$ are used only to check strong containment.

\section{Matching \emph{CGPs} via \emph{Strong Containment}}\label{section:matching_via_sc}


The main result of this Section is as follows:

\begin{theorem}\label{theorem:matching_via_sc}
For any two CGPs $C_{i\in[1,2]}$ and any data graph $G$, if $C_{_1}\sqsubseteq_s C_{_2}$ then $\mathcal{M}^{G}_{_{C_{_1}}}$ can be computed in $O(|C_{_1}|+|\mathcal{M}^{G}_{_{C_{_2}}}|)$ time without accessing $G$ at all.\hfill$\Square$
\end{theorem}

When a direct matching of $C_{_1}$ over $G$ is in $O(|C_{_1}|.|G|)$ time \cite{Mahfoud20,Mahfoud21}, \emph{strong containment} allows to reduce this time, if $C_{_1}\sqsubseteq_s C_{_2}$, by accessing only the match result of $C_{_2}$ over $G$ which is much smaller than $G$. We denote by $|\mathcal{M}^{G}_{_{C_{_2}}}|$ the number of data nodes and data edges in $G$ that match nodes and edges of $C_{_2}$ respectively. Indeed, $|\mathcal{M}^{G}_{_{C_{_2}}}|$ is at most $|V_{_2}|.|V|+|E_{_2}|.|E|$ but it is much smaller than $|G|$ (i.e. $|V|+|E|$) in practice.

One can provide a special matching algorithm to prove this Theorem. The idea of this algorithm is quite simple. Consider that $C_{_1}\sqsubseteq_s C_{_2}$ via the mapping $\lambda$. For any core node $u\in V_{_1}$, a set of its potential matches $S(u)$ is initialized to be $\bigcap_{u'\in\lambda(u)}\mathcal{M}^{G}_{_{C_{_2}}}(u')$. This set is refined later using the constraints in $C_{_1}$ (i.e. attributes constraints, labeling and \emph{CQs}) as well as the relations $\mathcal{R}^{+}$ and $\mathcal{R}^{-}$. For instance, for any $v\in S(u)$ and any negative predicate $p^{-}(u)\in\mathcal{P}^{-}_{_1}$ that is covered by $\mathcal{R}^{-}$ (i.e. evaluable by refinement), if there exists a subgraph $G_s(v)$ in $\mathcal{M}^{G}_{_{C_{_2}}}$ that matches $p^{-}(u)$ then $S(u)$ is refined by removing $v$ from it. Moreover, for any $v\in S(u)$ and any part of a positive predicate, let's be $p^{+}(u)$, that is covered by $\mathcal{R}^{+}$ (i.e. evaluable by refinement), if there is no subgraph $G_s(v)$ in $\mathcal{M}^{G}_{_{C_{_2}}}$ that matches $p^{+}(u)$ then $v$ is removed from $S(u)$. After the refinement process, if for some core node $u$ in $V_{_1}$, we get $S(u)=\emptyset$ then $C_{_1}\not\prec_{_c} G$ and $\mathcal{M}^{G}_{_{C_{_1}}}$ is empty. Otherwise, $\mathcal{M}^{G}_{_{C_{_1}}}$ can be extracted easily from the refined version of $S$.

\section{Conclusion}\label{section:conclusion}
We started by studying the application of the traditional containment for \emph{CGPs}, we discussed its complexity time, and we showed that its classical semantics allows to do \emph{sub-matching/over-matching} of \emph{CGPs} but not necessarily exact matching. To overcome this limit, we proposed \emph{strong containment} that takes in account the semantics of predicates and negation and allows to optimize well matching time of \emph{CGPs}. We showed that the new semantics does not come with a higher price since it is decidable in cubic time.
We are conducting an experimental study using real-life data graphs to check the effectiveness and efficiency of \emph{strong containment}. We are investigating the use of \emph{strong containment} for answering \emph{CGPs} using views. We notice that the presence of attributes and negation on nodes of \emph{CGPs} makes the problem very difficult compared to \cite{AnsweringGP_Using_Views} since the results of some \emph{CGP} can be obtained, not only by merging results of some views (as in \cite{AnsweringGP_Using_Views}), but also by applying some operations like intersection and difference. 

\bibliographystyle{splncs04}
\bibliography{WISE.bib}

\newpage
\appendix

\section*{Appendix}

\section{Proof of Lemma \ref{lemma:Complexity_of_QPoM}}
We prove the first part of Lemma \ref{lemma:Complexity_of_QPoM} by providing the algorithm  \textsf{POM} of Fig. \ref{Algorithm_POM} which inputs two \emph{QGPs} $Q_1$ and $Q_2$ and checks whether $Q_1\vartriangleright Q_2$, and if so, it returns the corresponding maximum match relation $S$. The algorithm simply follows from Def. \ref{definition:quantified_pattern_only_matching} and is easy to be understood, however, it does not run in quadratic time. Along the same lines as \cite{Henzinger, Mahfoud20}, a refinement of \textsf{POM} can be done to lead to a quadratic algorithm. The second part of Lemma \ref{lemma:Complexity_of_QPoM} can be proved by leveraging the result of \cite{Fan14} (see proof of Proposition 2.1).\hfill$\Square$

\begin{figure}[t!]
\rule{\linewidth}{0.5pt}
\scriptsize
\begin{flushleft}
\textbf{Algorithm} \textsf{POM}($Q_1,Q_2$)\newline
\textit{Input}: Two \emph{QGPs} $Q_{i\in[1,2]}$=($V_{_i},E_{_i},\mathcal{L}_{_i},\mathcal{A}_{_i},\mathcal{C}_{_i}$).\newline
\textit{Output}: The maximum match relation $S\subseteq V_{_1}\times V_{_2}$ if $Q_1\vartriangleright Q_2$, and $\emptyset$ otherwise.
\end{flushleft}
\begin{algorithmic}[1]
\State $S$ := $\{(u_1,u_2)\setminus u_1\in V_{_1},u_2\in V_{_2},\mathcal{L}_{_1}(u_1)=\mathcal{L}_{_2}(u_2),\mathcal{A}_{_1}(u_1)\sim\mathcal{A}_{_2}(u_2)\}$;
  
\Do
  \ForAll{$(u_1,u_2)\in S$ \textbf{and each} $e_2=(w_2,u_2)$ in $E_{_2}$}
      \If{($\nexists e_1=(w_1,u_1)\in E_{_1}$\textbf{:} $\mathcal{L}_{_1}(e_1)=\mathcal{L}_{_2}(e_2),(w_1,w_2)\!\in\!S$)}
	  \State $S$ := $S\setminus \{(u_1,u_2)\}$;
      \EndIf
  \EndFor
  \ForAll{$(u_1,u_2)\in S$ \textbf{and each} $e_2=(u_2,w_2)$ in $E_{_2}$}
      \If{($\nexists e_1=(u_1,w_1)\in E_{_1}$\textbf{:} $\mathcal{L}_{_1}(e_1)=\mathcal{L}_{_2}(e_2),(w_1,w_2)\!\in\!S,\mathcal{C}_{_1}(e_1)\!\geq\! \mathcal{C}_{_2}(e_2)$)}
	  \State $S$ := $S\setminus \{(u_1,u_2)\}$;
      \EndIf
  \EndFor
\doWhile{there are changes in $S$;}

\If{($\exists u_2\in V_{_2}$ \textbf{:} $\nexists u_1\in V_{_1}$ \textbf{with} $(u_1,u_2)\in S$)}
    \State \Return $\emptyset$;
\Else
    \State \Return $S$;
\EndIf

\normalsize
\end{algorithmic}

\rule{\linewidth}{0.5pt}
\caption{Algorithm for checking \emph{pattern-only matching} of \emph{QGPs}.}
\label{Algorithm_POM}
\end{figure}
\nointerlineskip

\section{Example of \emph{Pattern-Only Matching}}
Consider the \emph{CGPs} $C_1$ and $C_5$ of Example \ref{example:cgps} and their positive versions $C^{+}_1$ and $C^{+}_5$ respectively. It is easy to see that $C^{+}_5\vartriangleright C^{+}_1$, by matching the node \emph{Pr} (resp. \emph{PhD} and \emph{Article}) of $C^{+}_5$ to the node \emph{Pr} (resp. \emph{PhD} and \emph{Article}) of $C^{+}_1$. However, $C^{+}_1\not\vartriangleright C^{+}_5$ since: \textit{i}) the node \emph{Pr} of $C^{+}_1$ does not match the constraint ``\emph{@gender=female}'' defined over the node \emph{Pr} of of $C^{+}_5$; and \textit{ii}) no edge in $C^{+}_1$ can match the edge ${\scriptstyle PhD\xrightarrow[\geq 3]{member} Project}$ of $C^{+}_5$.\hfill$\Square$

\section{Proof of Theorem \ref{theorem:complexity_traditional_containment_CGPs}}
We prove Theorem \ref{theorem:complexity_traditional_containment_CGPs} by providing the algorithm \textsf{TContained} (shown in Fig. \ref{Algorithm_TContained}). Given two \emph{CGPs} $C_{_1}$ and $C_{_2}$ in input, it returns $\emptyset$ if $C_{_1}\not\sqsubseteq C_{_2}$, or the mapping $\lambda$ that allows $C_{_1}$ to be traditionally contained in $C_{_2}$ as well as the match relation $S$ that realizes $\lambda$. Notice that the relation $S$ is an intermediate data that allows to compute $\lambda$. However, we return it since both $\lambda$ and $S$ are necessary for strong containment checking. First of all, the algorithm checks whether $C^{+}_{_1}$ matches $C^{+}_{_2}$ (line 1) using a \emph{pattern-only-matching} algorithm \textsf{pom} (see Appendix). Next, the resulting match relation $S$ is refined by eliminating each pair ($u_1,u_2$) from it if: \textit{a}) the node $u_1$ of $C_{_1}$ does not satisfy negative predicates defined over $u_2$ of $C_{_2}$ (lines 2--4); or \textit{b}) $u_1$ is a core node while $u_2$ is a predicate node (lines 5--6), which is due to the fact that the traditional containment maps core nodes (resp. edges) of $C_{_1}$ to core nodes (resp. edges) of $C_{_2}$. The procedure \textsf{pom} is called over the refined version of $S$ to check whether this later still allows $C^{+}_{_1}$ to match $C^{+}_{_2}$ (line 7). If this is the case, then the set $\lambda(u)$ (resp. $\lambda(e)$) is computed based on $S$ for any core node $u$ (resp. core edge $e$) in $C_{_1}$ (lines 9--16). Finally, if some set $\lambda(u)$ (resp. $\lambda(e)$) is empty then $C_{_1}\not\sqsubseteq C_{_2}$ and the algorithm returns ($\emptyset,\emptyset$), otherwise, $C_{_1}\sqsubseteq C_{_2}$ and the pair ($\lambda,S$) is returned (lines 17--20).

Based on Lemma \ref{lemma:Complexity_of_QPoM}, it takes $O(|C^{+}_{_1}|.|C^{+}_{_2}|)$ time to check whether $C^{+}_{_1}$ matches $C^{+}_{_2}$ (line 1). Next, it takes $O(|\mathcal{P}^{-}_1|.|\mathcal{P}^{-}_2|)$ time to check whether, for each pair $(u_1,u_2)\in S$, $u_1$ satisfies all negative predicates of $u_2$ (lines 2--4). Moreover, it takes $O(|V_1|.|V_2|+|E_1|.|E_2|)$ time to compute the mapping $\lambda$ (lines 9--16). One can conclude that $|C_{_1}|=|C^{+}_{_1}|+|\mathcal{P}^{-}_1|$ (resp. $|C_{_2}|=|C^{+}_{_2}|+|\mathcal{P}^{-}_2|$). Therefore, the overall cost of algorithm \textsf{TContained} is bounded by $O(|C_{_1}|.|C_{_2}|)$ time, which completes the proof of Theorem \ref{theorem:complexity_traditional_containment_CGPs}.

\begin{figure}
\rule{\linewidth}{0.5pt}
\scriptsize
\begin{flushleft}
\textbf{Procedure} \textsf{ExtractRels}($C_{_1},C_{_2},S$)\newline
\textit{Input}: Two \emph{CGPs} $C_{i\in[1,2]}$=($V_{_i},E_{_i},\mathcal{L}_{_i},\mathcal{A}_{_i},\mathcal{C}_{_i},\mathcal{P}^{+}_{_i},\mathcal{P}^{-}_{_i}$), and a match relation $S$ that allows $C_{_1}\sqsubseteq C_{_2}$.\newline
\textit{Output}: The four relations $\mathcal{R}^{+}$, $\mathcal{R}^{-}$, $\mathcal{E}^{+}$ and $\mathcal{E}^{-}$.
\end{flushleft}
\begin{algorithmic}[1]
\State $\mathcal{R}^{+} := \emptyset$, $\mathcal{R}^{-} := \emptyset$, $\mathcal{E}^{+} := \emptyset$, $\mathcal{E}^{-} := \emptyset$;

\Statex \fbox{\textbf{\textit{Computing the relation $\mathcal{R}^{+}$}}}
\ForAll{$p^{+}_1(u)\in \mathcal{P}^{+}_{_1}$ \textbf{and each} $e_1=(u_1,w_1)$ in $p^{+}_1(u)$}
    \If{($\exists e_2=(u_2,w_2)$ in $E_{_2}$ \textbf{with} $\{(u_1,u_2),(w_1,w_2)\}\in S$)}
	  \State $\mathcal{R}^{+} := \mathcal{R}^{+} \cup \{(e_1,e_2)\}$;
    \EndIf
\EndFor

\Statex \fbox{\textbf{\textit{Computing the relation $\mathcal{E}^{+}$}}}
\State Compute $\overrightarrow{E_{_1}}(u)\subseteq E^{+}_{_1}$ and $\overleftarrow{E_{_1}}(u)\subseteq E^{+}_{_1}$ for each $u\in V^{+}_{_1}$;

\State Compute $\overrightarrow{E_{_2}}(w)\subseteq E^{+}_{_2}$ and $\overleftarrow{E_{_2}}(w)\subseteq E^{+}_{_2}$ for each $w\in V^{+}_{_2}$;

\ForAll{$p^{+}_2(w)\in \mathcal{P}^{+}_{_2}$ \textbf{and each} $(u,w)\in S$}
    \Statex \hspace{0.4cm}/* \textit{Find \textbf{E}dges in $\mathcal{P}^{+}_{_1}$ that are reachable from $u$ and \textbf{P}otentially \textbf{E}liminable over $p^{+}_2(w)$} */
    \State $PEE_u := \emptyset$;
    \State A queue \textsf{q} $:= \emptyset$; \textsf{q}.\textsf{push}($u,w$);
    \While{\textsf{q} $\neq\emptyset$}
        \State $(u,w) := \textsf{q}$.\textsf{pop}(); \textbf{mark}($u,w$);
        \ForAll{$e_2\!=\!(w,w')\in\overrightarrow{E_{_2}}(w)$ (resp. $e_2\!=\!(w',w)\in\overleftarrow{E_{_2}}(w)$)}
            \If{($\exists$ $e_1\!=\!(u,u')\in\overrightarrow{E_{_1}}(u)$ (resp. $e_1\!=\!(u',u)\in\overleftarrow{E_{_1}}(u)$) s.t: $e_1\equiv e_2$ via $S$)}
                \State $PEE_u := PEE_u\cup\{e_1\}$;
                \If{($(u',w')$ is not marked)}
                    \textsf{q}.\textsf{push}($u',w'$);
                \EndIf
            \EndIf
        \EndFor
    \EndWhile
    \Statex \hspace{0.4cm}\textbf{/* \textit{Refining} the set $PEE_u$ */}
    \State $V_u := \{u'~\backslash~ u'\neq u$ and $\exists e\in PEE_u$ s.t: $e$ starts/ends at $u'\}$;
    \State $adj(u') := \{u''~\backslash~ u''\in V_u$ and $(u',u'')$ (resp. $(u'',u')$) is in $PEE_u\}$;
    \ForAll{$u'\in V_u$}
        \If{($\exists$ an edge $e\in\overrightarrow{E}(u')\cup \overleftarrow{E}(u')$ s.t $e\not\in PEE_u$)}
            \State A queue \textsf{q} $:= \emptyset$; \textsf{q}.\textsf{push}($u'$);
             \While{\textsf{q} $\neq \emptyset$}
                 \State $u' :=$ \textsf{q}.\textsf{pop}(); \textbf{mark}($u'$);
                 \State $V_u := V_u~\backslash~\{u'\}$;
                 \ForAll{edge $e\!=\!(u',u'')$ (resp. $e\!=\!(u'',u')$) in $PEE_u$}
                    \State $PEE_u := PEE_u\backslash\{e\}$;
                     \If{($u''$ is not marked)}
                         \textsf{q}.\textsf{push}($u''$);
                     \EndIf
                 \EndFor
             \EndWhile
        \EndIf
    \EndFor
    \If{($PEE_u\neq\emptyset$)}
        \State Let $p^{+}_1(u)$ be a subgraph in $\mathcal{P}^{+}_{_1}$ composed by the nodes set $V_u$ and the edges set $PEE_u$;
        \State $\mathcal{E}^{+} := \mathcal{E}^{+} \cup \{(p^{+}_1(u),w)\}$;
    \EndIf
\EndFor

\Statex \fbox{\textbf{\textit{Computing the relation $\mathcal{R}^{-}$}}}
\ForAll{$p^{-}_{1}(u_1)\in \mathcal{P}^{-}_{_1}$ \textbf{and each} $(u_1,u_2)\in S$}
    \ForAll{undirected path $\textlangle e_1,\cdots,e_n\textrangle$ in $p^{-}_{1}(u_1)$ that starts at $u_1$}
        \ForAll{$(u_1,u_2)\in S$ \textbf{and} undirected path $\textlangle e'_1,\cdots,e'_n\textrangle$ in $C_{_2}$ that starts at $u_2$}
            \If{($e_i\sim e'_i$ for $1\leq i\leq n$)}
                \State $\mathcal{R}^{-} := \mathcal{R}^{-}\cup \{(e_i,e'_i)\}$;
            \EndIf
        \EndFor
    \EndFor
\EndFor

\Statex \fbox{\textbf{\textit{Computing the relation $\mathcal{E}^{-}$}}}
\ForAll{$p^{-}_{1}(u_1)\in \mathcal{P}^{-}_{_1}$ \textbf{and each} $(u_1,u_2)\in S$}
    \If{($\exists p^{-}_2(u_2)\in \mathcal{P}^{-}_{_2}$ \textbf{with} $p^{-}_{1}(u_1)\equiv p^{-}_2(u_2)$)}
	  \State $\mathcal{E}^{-} := \mathcal{E}^{-} \cup \{(p^{-}_{1}(u_1),u_2)\}$;
    \EndIf
\EndFor
\State\Return ($\mathcal{R}^{+},\mathcal{R}^{-},\mathcal{E}^{+},\mathcal{E}^{-}$);
\normalsize
\end{algorithmic}

\rule{\linewidth}{0.5pt}
\caption{Procedure to extract the relations $\mathcal{R}^{+}$, $\mathcal{R}^{-}$, $\mathcal{E}^{+}$ and $\mathcal{E}^{-}$.}
\label{Procedure_ExtractRels}
\end{figure}
\nointerlineskip

\section{Procedure \textsf{ExtractRels}}
The procedure \textsf{ExtractRels} is given in Fig. \ref{Procedure_ExtractRels}. Given two \emph{CGPs} $C_{_1}$ and $C_{_2}$ and a match relation $S$ that allows $C_{_1}$ to be traditionally contained in $C_{_2}$ (Lemma \ref{lemma:traditional_containment_of_cgps_nsc}). \textsf{ExtractRels} computes and returns the four relations $\mathcal{R}^{+}$, $\mathcal{R}^{-}$, $\mathcal{E}^{+}$ and $\mathcal{E}^{-}$. The computation of $\mathcal{R}^{+}$ (lines 2--4), $\mathcal{R}^{-}$ (lines 30--34) and $\mathcal{E}^{-}$ (lines 35--37) is quite simple and follows from Def. \ref{definition:refinement_elimination}. The computation of $\mathcal{E}^{+}$ is not a trivial task as we show hereafter. For each positive node $u$ in $\mathcal{P}^{+}_1$, we compute the sets $\overrightarrow{E_{_1}}(u)$ (resp. $\overleftarrow{E_{_1}}(u)$) of all positive edges that start (resp. end) at $u$ (line 5). Similar sets are computed for each positive node $w$ in $\mathcal{P}^{+}_2$ (line 6). For each positive predicate $p^{+}_2(w)$ in $\mathcal{P}^{+}_2$ and each pair $(u,w)$ in $S$, the goal is to check whether there exists a part of a positive predicate in $\mathcal{P}^{+}_1$, denoted by $p^{+}_1(u)$, that is centered at $u$ and eliminable over $w$: i.e. there exists a part of $p^{+}_2(w)$ that is equivalent to $p^{+}_1(u)$. To check that, the procedure firstly computes a set $PEE_u$ that includes all edges in $\mathcal{P}^{+}_1$ that are reachable from $u$ and equivalent to some edges in $p^{+}_2(w)$ via $S$. A \emph{Breadth-First Search} \cite{Cormen} is applied (lines 8--15) to compute the set $PEE_u$. The edges in $PEE_u$ are concerned by an elimination over $w$ only if they satisfy condition (2b) of Def. \ref{definition:refinement_elimination}. Thus, the procedure refines the set $PEE(u)$ (lines 16--26) by eliminating all edges that do not satisfy this condition. This refinement is done in terms of nodes rather than edges in order to reduce the time complexity. In a nutshell, the set $V_u$ is extracted from $PEE_u$ to include all positive nodes in $\mathcal{P}^{+}_1$ that play a role in $PEE_u$ (line 16). Next, for each node $u'$ in $V_u$, if some adjacent edges to $u'$ in $\mathcal{P}^{+}_1$ do not belong to $PEE_u$ (line 19), then all edges that start/end at $u'$ in $PEE_u$ do not satisfy the aforementioned condition (2b) and must be eliminated from $PEE_u$. Once eliminated from $PEE_u$, their adjacent edges will no longer satisfy condition (2b) and so on. For this reason, all edges in $PEE_u$ that connect $u$ to $u'$ are eliminated from $PEE_u$ in a recursive manner (lines 20--26).
Finally, if the refined version of $PEE_u$ is not empty then there exists a part of a positive predicate in $\mathcal{P}^{+}_1$, denoted by $p^{+}_1(u)$, that is eliminable over $w$ and induced by the nodes set $V_u$ and the edges set $PEE_u$ (lines 27--28). Hence, the pair $(p^{+}_1(u),w)$ is added to the relation $\mathcal{E}^{+}$ (line 29).

\section{Complexity Analysis of \textsf{SContained}}
Consider first the procedure \textsf{ExtractRels}. On can verify that the relation $\mathcal{R}{+}$ can be computed in $O(|E^{+}_{_1}|.|E_{_2}|)$ time (lines 2--4), while $\mathcal{R}{-}$ can be computed in $O(|E^{-}_{_1}|.|E_{_2}|)$ time (lines 30--34). Since the equivalence between two \emph{QGPs} can be checked via a bidirectional containment, then for each pair of negative predicates $p^{-}_{1}(u_1)$ and $p^{-}_2(u_2)$, it takes $O(|p^{-}_{1}(u_1)|.|p^{-}_2(u_2)|)$ time (from Theorem \ref{theorem:complexity_traditional_containment_CGPs}) to check whether $p^{-}_{1}(u_1)\equiv p^{-}_2(u_2)$. Thus, the relation $\mathcal{E}^{-}$ can be computed in $O(|\mathcal{P}^{-}_{_1}|.|\mathcal{P}^{-}_{_2}|)$ time by considering all possible pairs (lines 35--37). The cost required to compute $\mathcal{E}^{+}$ is detailed as follows. The different sets $\overrightarrow{E_{_1}}$ and $\overleftarrow{E_{_1}}$ are computed for all positive nodes of $\mathcal{P}^{+}_{_1}$ in $O(|V^{+}_{_1}|+|E^{+}_{_1}|)$ time (line 5), which is equivalent to $O(|\mathcal{P}^{+}_{_1}|)$. Similarly, the different sets $\overrightarrow{E_{_2}}$ and $\overleftarrow{E_{_2}}$ are computed for all positive nodes of $\mathcal{P}^{+}_{_2}$ in $O(|\mathcal{P}^{+}_{_2}|)$ time (line 6). 
For each positive predicate $p^{+}_2(w)$ in $\mathcal{P}^{+}_{_2}$, let $E_w$ (resp. $V_w$) be the edges set (resp. nodes set) of this predicate. For each node $u\in V^{+}_{_1}$, the set $PEE_u$ can be computed in $O(|E^{+}_{_1}|.|E_w|+|V^{+}_{_1}|.|V_w|)$ time (lines 8--15). This set is refined later (lines 16--26) in $O(|V^{+}_{_1}|+|E^{+}_{_1}|)$ time, i.e. in $O(|\mathcal{P}^{+}_{_1}|)$ time. If the refined version of $PEE_u$ is not empty, then the corresponding subpredicate $p^{+}_{1}(u)$ is computed and added to $\mathcal{E}^{+}$ in $O(|\mathcal{P}^{+}_{_1}|)$ time (lines 27--29). By considering all positive predicates in $p^{+}_2(w)$ in $\mathcal{P}^{+}_{_2}$ and all possible pairs $(u,w)\in S$, all possible sets $PEE_u$ can be computed in $O(|V^{+}_{_1}|.(|E^{+}_{_1}|.|E^{+}_{_2}|+|V^{+}_{_1}|.|V^{+}_{_2}|))$ time and refined in $O(|V^{+}_{_1}|.|V^{+}_{_2}|.|\mathcal{P}^{+}_{_1}|)$ time. Therefore, the overall cost related to the computation of the relation $\mathcal{E}^{+}$ is bounded by $O(|V^{+}_{_1}|.|\mathcal{P}^{+}_{_1}|.|\mathcal{P}^{+}_{_2}|)$.

Given the above, the overall cost of procedure \textsf{ExtractRels} is bounded by $O(|E^{+}_{_1}|.|E_{_2}|+|E^{-}_{_1}|.|E_{_2}|+|\mathcal{P}^{-}_{_1}|.|\mathcal{P}^{-}_{_2}|+|V^{+}_{_1}|.|\mathcal{P}^{+}_{_1}|.|\mathcal{P}^{+}_{_2}|)$. Notice that the cost $O(|\mathcal{P}^{-}_{_1}|.|\mathcal{P}^{-}_{_2}|)$ time (resp. $O(|E^{+}_{_1}|.|E_{_2}|+|V^{+}_{_1}|.|\mathcal{P}^{+}_{_1}|.|\mathcal{P}^{+}_{_2}|)$ time) is due to the evaluability checking of negative (resp. positive) predicates of $C_{_1}$ over $C_{_2}$.

The complexity of algorithm \textsf{SContained} is stated as follows. From Theorem \ref{theorem:complexity_traditional_containment_CGPs}, it requires $O(|C_{_1}|.|C_{_2}|)$ time to check whether $C_{_1}\sqsubseteq C_{_2}$ (line 1). Next, the aforementioned cost of procedure \textsf{ExtractRels} is required to compute the relations $\mathcal{R}^{+}$, $\mathcal{R}^{-}$, $\mathcal{E}^{+}$ and $\mathcal{E}^{-}$ (line 3). The set $CE^{+}$ is computed in $O(|E^{+}_{_1}|.|E_{_2}|+|V^{+}_{_1}|.|V^{+}_{_2}|.|E^{+}_{_1}|)$ time (lines 5--6) while the set $CE^{-}$ takes $O(|E^{-}_{_1}|.|E_{_2}|+|V^{+}_{_2}|.|E^{-}_{_1}|)$ time (lines 7--8). Hence, the overall cost of \textsf{SContained} is bounded by $O(|C_{_1}|.|C_{_2}|+|V^{+}_{_1}|.|\mathcal{P}^{+}_{_1}|.|\mathcal{P}^{+}_{_2}|)$ time, which completes the proof of Theorem \ref{theorem:complexity_strong_containment_CGPs}.

\end{document}